\documentclass[preprint]{revtex4}

\usepackage{graphicx}
\usepackage{dcolumn}
\usepackage{bm}

\usepackage{amsmath}
\usepackage{amssymb}

\newcommand{\be}{\begin{eqnarray}}
\newcommand{\ee}{\end{eqnarray}}

\newcommand \beq{\begin{eqnarray}}
\newcommand \eeq{\end{eqnarray}}

\def\0{\over } \def\2{{1\over2}} \def\4{{1\over4}}
\def\5{\hat } \def\6{\partial }

 \def\b{\beta }  \def\d{\delta }

\def\({\left(} \def\){\right)} \def\<{\langle } \def\>{\rangle }
\newcommand{\nn}{\nonumber\\ }

\input epsf

\def\del{\partial}

\def\frac#1#2{{#1 \over #2}}

\def\half{\ifinner {\scriptstyle {1 \over 2}}
    \else {1 \over 2} \fi}

\def\simge{\mathrel{%
    \rlap{\raise 0.511ex \hbox{$>$}}{\lower 0.511ex \hbox{$\sim$}}}}
\def\simle{\mathrel{
    \rlap{\raise 0.511ex \hbox{$<$}}{\lower 0.511ex \hbox{$\sim$}}}}

\def\slashchar#1{\setbox0=\hbox{$#1$}
    \dimen0=\wd0
    \setbox1=\hbox{/} \dimen1=\wd1
    \ifdim\dimen0>\dimen1
       \rlap{\hbox to \dimen0{\hfil/\hfil}}
       #1
    \else
       \rlap{\hbox to \dimen1{\hfil$#1$\hfil}}
       /
    \fi}

 \def\b{\beta }  \def\d{\delta }

\def\0{\over } \def\2{{1\over2}} \def\4{{1\over4}}
\def\5{\hat } \def\6{\partial }

\def\Tr{{\,\mathrm{ Tr}\,}}

\def\del{\partial }

\newcommand{\bea}{\begin{eqnarray}}
\newcommand{\eea}{\end{eqnarray}}

\begin{document}

\preprint{SPhT-T03/187}
\vspace*{.5cm}

\title{Renormalization of $\Phi$-derivable approximations in scalar 
field theories}

\author{Jean-Paul Blaizot, Edmond Iancu and Urko Reinosa}
\email{blaizot@spht.saclay.cea.fr, iancu@spht.saclay.cea.fr,
reinosa@hep.itp.tuwien.ac.at}
  \affiliation{Service de Physique Th\'eorique, CEA/DSM/SPhT,
91191 Gif-sur-Yvette Cedex, France.}

\date{\today}

\vspace*{1.cm}
\begin{abstract}
We discuss the renormalization of $\Phi$-derivable approximations for 
scalar field theories. In such approximations, the self-energy is 
obtained as the solution of a self-consistent equation which 
effectively resums infinite subsets of  diagrams of perturbation theory. 
We show that a consistent renormalization can be carried out, 
and we provide an explicit construction of the counterterms 
needed to eliminate the subdivergences. The counterterms are calculated
from the solution of an auxiliary gap equation which determines 
the leading asymptotic part of the self-energy. This auxiliary gap equation 
may be chosen as the gap equation of the massless theory at zero temperature. 
We verify explicitly that the counterterms determined at zero temperature  
are sufficient to eliminate the divergences which occur in finite temperature 
calculations. 
\end{abstract}

\pacs{Valid PACS appear here}
\maketitle

\section{Introduction }
\setcounter{equation}{0}

Self-consistent 
approximations, such as the $\Phi$-derivable approximations, offer a way to deal with physical
situations where the quasiparticle picture is a good starting point. They allow one to include
large effects of the interactions into the propagators of quasiparticles, leaving relatively weak residual
interactions among them.  Such approximations were introduced many  years ago in the
context of the non-relativistic many body problem
\cite{Luttinger:1960,Baym,DeDominicis:1964}, and have been reformulated later 
in the language of relativistic field theory 
\cite{Cornwall:vz}. They  have been  used recently in various
   field theoretical problems. In particular, they form the basis for a
quasiparticle description of  the  equilibrium thermodynamics of the quark-gluon plasma
\cite{Blaizot:1999ip}, which exploits 
the remarkable simplifications they lead to in the  calculation of the entropy
\cite{Vanderheyden98}
(along similar lines, models inspired by the 2-loop $\Phi$-derivable
approximation for QCD have also been derived \cite{Peshier:2000hx}).  
They constitute a framework for 
 variational approximations \cite{Braaten:2001en} 
and they provide a more general  alternative to 
 screened perturbation theory
\cite{Karsch:1997gj}. They have  been recently applied to   
the dynamics of  quantum fields out of equilibrium
\cite{Aarts:2002dj}. $\Phi$-derivable approximations are known to be ``conserving'' approximations \cite{Baym}, that is,
they are consistent with global symmetries of the Lagrangian \cite{vanHees:2002bv}. However since
they lead to modifications of  the propagators while leaving the vertices unaffected, they violate
the Ward identities associated to local symmetries; consequences of such violations have  been recently explored 
quantitatively
\cite{Arrizabalaga:2002hn}, and shown to be suppressed with respect to
naive estimates based on power counting.

One of the main obstacles
for implementing  such non-perturbative approximations in quantum
field theory is the difficulty of their renormalization. The central equation that one has to
solve is a self-consistent Dyson equation, 
or gap equation, for the propagator. Such a gap
equation  effectively resums an infinite set  of Feynman diagrams contributing to the self-energy,
with arbitrary powers in the coupling constant. We loose the expansion in powers in the 
coupling as a tool to organize the
divergences of the various diagrams. Indeed, at each order in an expansion of the coupling, only a
subset of the diagrams of perturbation theory contributing to this order is effectively taken into
account by the gap equation. The standard proofs of renormalizability
 do not therefore
immediately guarantee that the divergences can be systematically eliminated.  Thus, with the
exception of the  simplest
self-consistent mean field approximations, which can be solved easily (see for instance
\cite{Dolan:qd,Drummond:1997cw}), the renormalization of $\Phi$-derivable approximations
has remained a difficult and unsolved problem.

One of the motivations for studying $\Phi$-derivable approximations 
is their potential usefulness in quantum
field theory at finite temperature. A problem that one meets there 
is the possible occurence of temperature dependent infinities.  While, on general  grounds
(see for instance \cite{Collins:xc}),  one does not expect the temperature to generate any new infinities, it is
not always straightforward to guarantee  in practical calculations beyond perturbation theory that potentially
divergent terms dependent on the temperature do 
actually cancel. This is especially the case in situations where
the various divergences  cannot be classified according to some expansion in a small parameter. (For a simple
model calculation where one can verify  this cancellation explicitly in the first non trivial order in a 
$1/N$ expansion, see
\cite{Blaizot:2002nh}.) It may happen then that one is  led to 
introduce counterterms which depend on the
temperature  (see for instance \cite{Braaten:2001en}), 
which obscures the physical interpretation of the results.

 A major progress on this issue was achieved recently by  van Hees
and Knoll in a series of papers \cite{vanHees:2001ik,VanHees:2001pf,vanHees:2002bv}.  The strategy 
that they put forward
\cite{vanHees:2001ik} is based on an expansion of the 
 self-consistent propagator around the corresponding propagator in
the vacuum. They were
able to prove the elimination of the temperature 
dependent divergences, and have established a well
defined scheme to solve the gap equation 
\cite{vanHees:2002bv,VanHees:2001pf}. 
However, several features of their approach
remain somewhat unsatisfactory.  The  real time formalism that they use, 
which appears to be essential at some stages of their derivation,
 does not allow an easy comparison with
more conventional field  theoretical techniques. More importantly, 
by focusing on the temperature--dependent divergences alone,
they introduce   disymmetrical
treatments of the vacuum sector and the finite  temperature one. 
This hides the fact that the particular 
structure of subdivergences that one needs to deal with to guarantee 
the elimination of temperature dependent 
infinities is already present in the vacuum. 
Thus, they do not address the issue of renormalizability 
in the vacuum sector itself, but take it for granted.
Finally, the approach  by  van Hees and Knoll cannot be applied
as it stands to massless theories, since an expansion around the vacuum
(massless) propagator would be afflicted with infrared divergences.
This spoils a main benefit of
the self-consistent approximations, which is to regulate the
infrared behaviour via a resummation of the medium effects.

In this paper,  we present a general discussion of the
renormalization of
$\Phi$-derivable approximations for scalar field theories. 
We do not rely  on a separation of the vacuum sector  from
the finite temperature one, and shall  
use the imaginary time formalism, making the connection
with  standard field theory transparent. 
In fact we carry out the renormalization at zero temperature, and
proceed as in ordinary perturbation theory, by analyzing the 
subdivergences of the diagrams which are generated
as one iterates the gap equation. 
The analysis and elimination of divergences is carried out according to the 
Bogoliubov-Parasiuk-Hepp (BPH) procedure \cite{BPH,Collins:xc}, 
by first removing subdivergences using previously 
determined counterterms and then absorbing the remaining divergences 
into corrections to these counterterms. As we shall see, 
vertex divergences occur, which are identified with those of
a special Bethe-Salpeter equation for the 4-point function. 
The counterterms will be explicitly constructed, as
well as a finite gap equation. At the end of our analyzis, 
it will become straightforward to show that the
temperature--dependent contributions are finite, as expected: 
potential temperature dependent divergences cancel
out as one eliminates the subdivergences. But since carried out
already in the vacuum, our proof of the renormalizability
of the $\Phi$-derivable approximations is a general one, 
and applies to any of their potential applications,
in particular to off--equilibrium problems.

The outline of the paper is the following. 
In the next section, we recall the essential features
of $\Phi$-derivable approximations for the $\phi^4$ scalar field
theory, their diagrammatic content and basic definitions concerning their renormalization. In the
next two sections,  we  discuss thoroughly the renormalization of
simple approximations based on the 2-loop and 3-loop skeletons, respectively. The discussion is
carried out in such a way as to prepare for the  generalization presented in the last section.  A
preliminary account of this work was given in
\cite{Blaizot:2003br}.

\section{General formalism\label{sec:formalism}}

As a generic example of a scalar field theory, we shall consider that described by the Lagrangian
\beq
{\cal L}=\frac{1}{2}\left(\del\phi\right)^2-\frac{ m^2}{2}\phi^2-\frac{\lambda}{4!}\phi^4, 
\eeq
and  discuss only the symmetric phase in which the field expectation
value vanishes, that is,  we shall not treat the case of spontaneously broken symmetries.

\subsection{Skeleton expansion for the thermodynamical potential}

The starting point of $\Phi$-derivable approximations is the following expression for 
the thermodynamic potential $\Omega$, considered as a  functional of the full propagator $D$
\cite{Luttinger:1960,DeDominicis:1964,Baym}:
\be
\label{LW}\b \Omega[D]=\2 \Tr \log D^{-1} -\2 \Tr \Pi D+\Phi[D]\,,
\ee 
where $\Tr$ denotes the trace in configuration space,
and $\b=1/T$, with $T$ the temperature. The trace  over configuration space  involves integration over
imaginary time and over spatial coordinates. Alternatively, these can be
turned into summations over Matsubara frequencies and integrations over
spatial momenta \cite{Kap:FTFT,LeB:TFT}:
\beq
\int_0^\beta {\rm d}\tau
\int {\rm d}^3x \rightarrow \beta V
 \int [{\rm d}k] ,\eeq
where $V$ is the spatial volume,
 $K^\mu=(i\omega_n, {\bf k})$
and $\omega_n = n\pi T$, with $n$ an even integer.
We have introduced a condensed notation for 
the measure of the loop integrals (i.e., the sum over the
Matsubara frequencies $\omega_n$ and the integral over the
spatial momentum ${\bf k}$): 
\beq\label{mesure}
\int[{\rm d}k]\equiv T\sum_{n, even} \int\frac{{\rm d}^3k}{(2\pi)^3}\,.
\eeq
At zero temperature, $\beta\to\infty$, the loop integrals become
 ordinary Euclidean integrals in four dimensions:
\beq\label{mesure2}
\int[{\rm d}k]\longrightarrow \int\frac{{\rm d}^4k}{(2\pi)^4},
\eeq
and $\Omega$ becomes the ground state energy. In the following we shall often  use the short-hand notation
$\int_K$ to denote either the sum integral (\ref{mesure}) at finite temperature, or the Euclidean
integral (\ref{mesure2}), depending upon the context.

The sum-integrals in equations like  Eq.~(\ref{LW})
contain  ultraviolet divergences  and  require  regularization. Unless explicitly stated otherwise, we
shall assume throughout dimensional regularization.

\begin{figure}[htb]
\epsfysize=4cm
\centerline{\epsffile{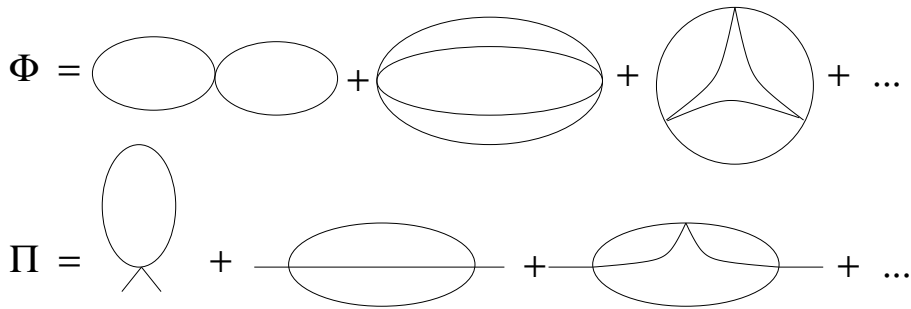}}
\caption[a]{The skeleton diagrams contributing to the thermodynamical potential to order four
loops in scalar $\phi^4$ theory, and, below, the corresponding contributions to the self energy to
order three loops.
\label{skeletons}}
\end{figure}

The self-energy $\Pi$ in  Eq.~(\ref{LW}) is related to $D$ by
Dyson's equation
\beq\label{Dyson}
D^{-1}=D^{-1}_0+\Pi, 
\eeq
where  $D_0$ denotes the free propagator.  

The quantity $\Phi[D]$ is the sum of the 2-particle-irreducible ``skeleton''
diagrams, i.e., diagrams  that cannot be separated into two
disconnected parts by cutting two lines. In short, skeletons are diagrams of perturbation
theory in which one cannot identify self-energy insertions. The skeletons contributing to
the thermodynamic potential up to four loops and, respectively,  to the self-energy up to three loops, are
displayed
 in Fig.~\ref{skeletons}. In Eq.~(\ref{LW}), $\Phi[D]$ is
calculated by associating full propagators $D$ to the lines of the skeletons (while
skeleton diagrams in perturbation theory would be calculated with free propagators $D_0$).

There are several ways to obtain Eq.~(\ref{LW}): by integration over the coupling constant \cite{Luttinger:1960}, 
by a Legendre transform with respect to a bilocal source
\cite{DeDominicis:1964,Cornwall:vz}, or via a simple diagrammatic analysis
\cite{DeDominicis:1964}.  We briefly recall here a derivation based on the latter approach. This
will give us the oportunity to introduce  concepts that will be useful later on. 

Consider then the
self-energy
$\Pi$. From its definition in Eq.~(\ref{Dyson}),  it is the 
sum of all the one-line-irreducible diagrams contributing to the
propagator. Clearly, the  self-energy skeletons can be obtained from the 
skeletons of $\Omega$ by cutting one line of the latter (see Fig.~\ref{skeletons}  for examples). 
Thus we have:
\be\label{PhiPi1}
\frac{\delta \Phi[D_0]}{\delta D_0}={1\over 2}\Pi_s[D_0],
\ee
where $\Pi_s[D_0]$ denotes the sum of self-energy skeletons evaluated with free
propagators. Now, it is easy to
verify that all the diagrams of perturbation theory contributing to $\Pi$ can be obtained
from the  self-energy skeletons by replacing their internal lines  by full propagators.  In other
words:
\be\label{PhiPi2}
\Pi=\Pi_s[D]=2\frac{\d\Phi[D]}{\d D}.
\ee

\begin{figure}[htb]
\epsfysize=4.cm
\centerline{\epsffile{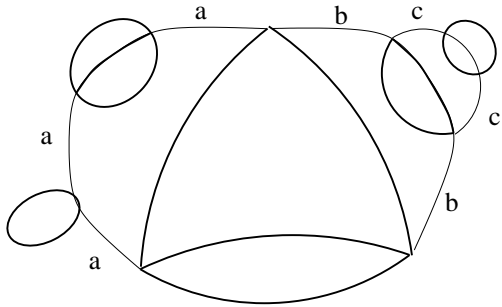}}
\vspace{3mm}
\caption[a]{A typical diagram contributing to the 
thermodynamical potential\label{typical}}
\end{figure}

This simple procedure  does not hold, however, for the diagrams contributing to the
thermodynamical potential because, in the absence of external lines, the  diagrams obtained by replacing
the free propagators in a  skeleton by full propagators are overcounted.  
Consider, as an example, the
diagram in Fig.~\ref{typical}.  One can identify on it five $\Phi$-skeletons 
(by replacing successively by
propagators the parts which can be regarded as self-energy insertions); 
these skeletons belong to the set 
drawn in the first line of Fig.~\ref{skeletons}. 
The lines labelled $a$, $b$, and $c$ in Fig.~\ref{typical} are said to  belong to
\textit{cycles} (there are three cycles in Fig.~\ref{typical}). 
The cycles have the following property: By cutting two lines of a
cycle, one cuts the diagram into two disconnected parts. 
Cutting two lines belonging to two
different cycles leaves the diagram connected. 
Clearly, two cycles of lines have in common at most 
one $\Phi$-skeleton (identified by replacing each cycle 
 attached to it by a propagator), 
but a skeleton may have several cycles attached to it.  
In other words the cycles generate a tree
structure, whose branches are chains of skeletons attached 
by cycles of lines (an illustration of such a structure  
is given in Fig.~\ref{typicalPi} below). 
This particular structure underlies the algebraic derivation of Eq.~(\ref{LW})
in terms of a Legendre transform with respect to a bilocal source \cite{DeDominicis:1964,Cornwall:vz}.

For a given diagram $\gamma$, let $n_c$ be the number of cycles,
$n_s$ the number of $\Phi$-skeletons and $n_l$ the number 
of lines belonging to cycles. The following
identity is easily proved by induction:
\beq
n_c+n_s-n_l=1.
\eeq
For the diagram in Fig.~\ref{typical}, we have $n_c=3$, $n_s=5$, $n_l=7$. Let now $w(\gamma)$ be the
contribution of the diagramm $\gamma$, calculated with the rules of perturbation theory
(i.e., with free propagators). We have  the obvious identity:
\beq
\sum_\gamma w(\gamma)=\sum_\gamma\left[n_c(\gamma)+n_s(\gamma)-n_l(\gamma)
\right]w(\gamma).
\eeq
The usefulness of this relation is that the three terms in its right hand side correspond  to
the three terms in Eq.~(\ref{LW}), as we now show \cite{BR86}. 

We have :
\beq\label{LW1}
\Phi[D]=\sum_\gamma n_s(\gamma) w(\gamma).
\eeq
This states simply that when one evaluates the skeletons with full propagators one counts
each diagram $\gamma$ as many times as one can identify skeletons in $\gamma$. For instance the
diagram of Fig.~\ref{typical} will occur once in the 1-loop skeleton of Fig.~\ref{skeletons}, three times in the
 2-loop skeleton, and once in the 3-loop skeleton. 

\begin{figure}[htb]
\epsfysize=4.cm
\centerline{\epsffile{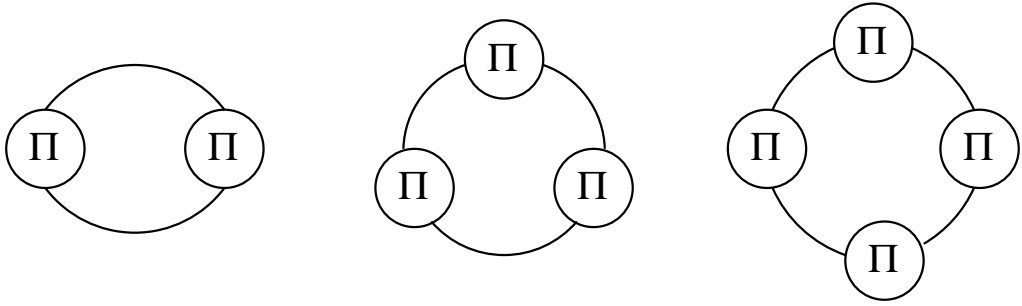}}
\vspace{3mm}
\caption[a]{Ring diagrams, whose contribution is given in Eq.~(\ref{LW2})\label{Ring}}
\end{figure}

Consider now the sum of the family of  diagrams of Fig.~\ref{Ring} corresponding to a cycle of
lines with an arbitrary number $n$ ($n\ge 2$) of self-energy insertions. This is given by 
\beq\label{LW2}
\2 \Tr \log (1+D_0\Pi)-\2\Tr D_0\Pi=\sum_\gamma n_c(\gamma) w(\gamma).
\eeq
Again, the right hand side follows from the fact that the diagram $\gamma$ is counted as many
times in the left hand side as there are cycles of lines in $\gamma$.

Finally consider one line of a cycle, and view it as part of a full propagator  closed
on a self-energy. We have then
\beq\label{LW3}
-\2 \Tr \Pi D+\2 \Tr \Pi D_0=\sum_\gamma n_l(\gamma) w(\gamma),
\eeq
since $\gamma$ will be counted as many times in the left hand side as it contains lines of
cycles. 

By adding together the contributions obtained in Eqs.~(\ref{LW1}), (\ref{LW2}) and
(\ref{LW3}) we recover Eq.~(\ref{LW}) up to the term term $(1/2) \Tr \log D_0^{-1}$ which is proportional
to the thermodynamical potential in the absence of interactions.

\subsection{$\Phi$-derivable approximations}

An important property of the functional $\Omega[D]$ is to be
stationary under variations of $D$ (at fixed $D_0$) around the
physical propagator: 
\be\label{selfcons}
{\d\Omega[D] / \d D}=0. 
\ee
It is indeed easily verified that this equation for $D$ is equivalent to the self-consistent Dyson equation
(commonly
referred to as the gap equation):
\beq\label{Dysonsc}
D^{-1}=D^{-1}_0+\Pi[D],\qquad \Pi[D]=2\frac{\delta \Phi[D]}{\delta D}\,.
\eeq
The physical thermodynamical potential is then obtained as the value of $\Omega[D]$ for $D$ solution of this equation.

Self-consistent (or variational)
approximations, i.e., approximations which preserve the
stationarity property (\ref{selfcons}),
are  obtained by selecting a class of skeletons in
$\Phi[D]$ and calculating $\Pi$ from Eq.~(\ref{PhiPi2}).
Such approximations are commonly called ``$\Phi$-derivable''
\cite{Baym}. 

\begin{figure}[htb]
\epsfysize=7cm
\centerline{\epsffile{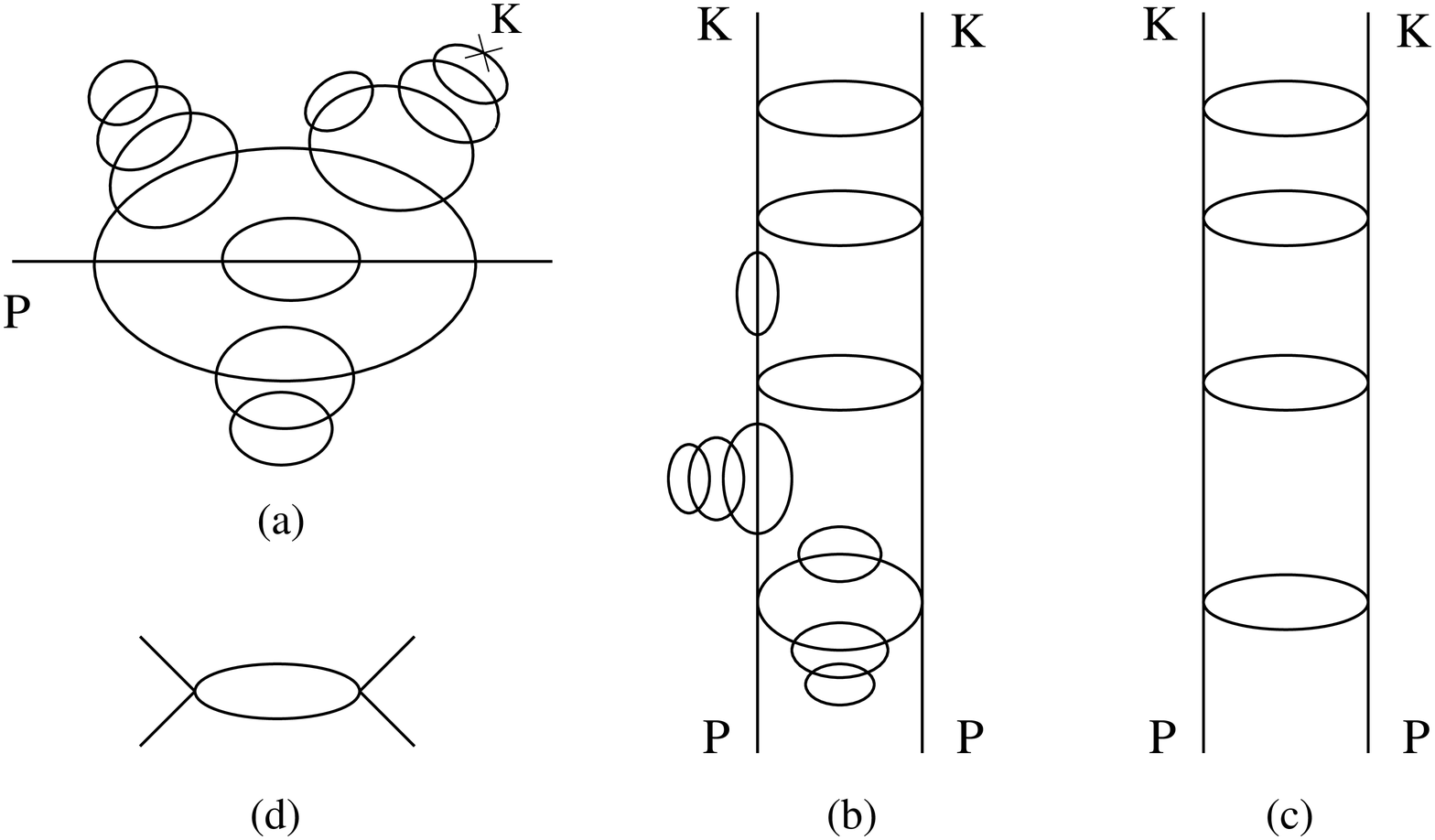}}
\vspace{3mm}
\caption[a]{ (a): A typical diagram contributing 
to $\Pi$ when the 3-loop skeleton is chosen for $\Phi$. (b): A diagram
contributing to $\Gamma(P,K)$, and which is obtained by opening 
the line marked by a cross in (a). (c): The corresponding
contribution to $\Gamma_0(P,K)$. (d): The kernel $\Lambda_0$ whose iteration
 produces the diagram (c).}
\label{typicalPi}
\end{figure}

By selecting a skeleton in $\Phi$, one obtains an approximation that effectively resums, through the iteration of the
gap equation, an infinite subset of diagrams of perturbation theory. An example of such a resummation is illustrated in
Fig.~\ref{typicalPi}.a corresponding to the approximation based on the 3-loop skeleton. One recognizes the topology
discussed above for  the diagram contributing to $\Omega$: a tree
structure, 
with skeletons joined together by cycles of lines. Consider now the diagrams
obtained by opening one line of
$\Pi(P)$ which does not belong to a cycle of
lines (thereby selecting one branch in the tree alluded to before). Let $K$ be the momentum flowing
through the corresponding propagator $D_0$. The resulting diagram is a contribution to a 4-point function that we shall call $\Gamma(P,K)$ (see Fig.~\ref{typicalPi}.b). 
The diagram may contain cycles with more than two lines, that is, lines carrying self-energy
insertions. Let  then $\Gamma_0$
be the set of diagrams of $\Gamma$ which do not carry such insertions, i.e.,  
each cycle of $\Gamma_0$ contains only two lines (see Fig.~\ref{typicalPi}.c). 
The diagrams of $\Gamma_0$ are  two-line reducible. Let us call
$\Lambda_0$ the contribution to $\Gamma_0$ which is two-line irreducible (cf.  
Fig.~\ref{typicalPi}.d): $\Lambda_0(P,K)$ is the sum of all
the diagrams contributing to $\Gamma_0(P,K)$ 
which, by cutting two internal lines, cannot be split into
two pieces, one containing the lines carrying $P$ the other the lines
carrying $K$.  Clearly,  the diagrams contributing to $\Gamma_0$ have the form of chains of
$\Lambda_0$'s joined together by pairs of propagators $D_0$ (see Fig.~\ref{typicalPi}.c).  The
diagrams that we have identified play the role of skeletons for $\Gamma$ and $\Lambda$,
respectively. As was the case for $\Pi$, the presence of external lines makes the identification
of skeletons unambiguous, and the complete set of diagrams contributing to
$\Gamma$ and $\Lambda$ can be obtained  by replacing in $\Gamma_0$ and $\Lambda_0$ the free
propagators
$D_0$ by the full propagator $D$ (cf. Eq.~(\ref{PhiPi2}) for $\Pi$). 

\begin{figure}[htb]
\epsfysize=2.5cm
\centerline{\epsffile{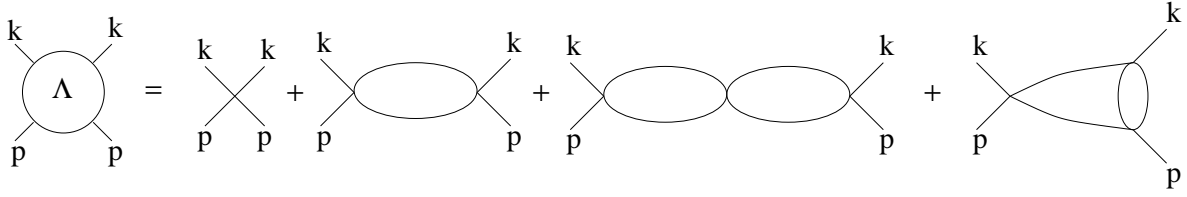}}
\caption[a]{The skeleton diagrams contributing to the
kernel of the Bethe-Salpeter equation up two loops \label{kernelBSLambda}}
\end{figure}

\begin{figure}[htb]
\epsfysize=4cm
\centerline{\epsffile{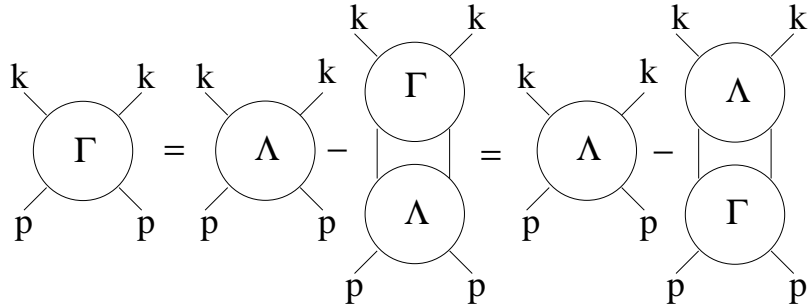}}
\caption[a]{The Bethe-Salpeter equation in the $t$-channel\label{tchannelBS}}
\end{figure}

It it not difficult to show  that  $\Lambda$ can be obtained from $\Pi$ by functional differentiation:
\beq\label{Lambdadef}
\Lambda(P,K)=2\frac{\delta \Pi(P)}{\delta D(K)}=4\frac{\delta^2 \Phi}{\delta D(K)\delta D(P)}=\Lambda(K,P).
\eeq
The contributions to $\Lambda$, up to 2-loop, are displayed in Fig.~\ref{kernelBSLambda}. 
As for the diagrams of $\Gamma$, they  are 
obtained by solving iteratively the following  Bethe-Salpeter equation (see Fig.~\ref{tchannelBS}):
\beq\label{BSdef}
\Gamma(P,K)&=&\Lambda(P,K)-\frac{1}{2}\int_Q \Lambda(P,Q)D^2(Q)\Gamma(Q,K)\nonumber\\
&=&\Lambda(P,K)-\frac{1}{2}\int_Q \Gamma(P,Q)D^2(Q)\Lambda(Q,K).
\eeq
 Note that the symmetry
property $\Lambda(P,K)=\Lambda(K,P)$ entails the corresponding property for the full 4-point function:
$\Gamma(P,K)=\Gamma(K,P)$. This is easily verified by solving Eq.~(\ref{BSdef}) iteratively. Note also
that the Bethe-Salpeter equation (\ref{BSdef}) realizes a resummation in a specific channel, which one may
call the $t$-channel: this is the channel in which the kernel $\Lambda$ is 2-particle
 irreducible.
It follows that as soon as one restricts oneself to a limited class of skeletons, one looses
the crossing symmetry of the general 4-point function. This is a general difficulty with
$\Phi$-derivable approximations (see also \cite{vanHees:2002bv}).

Although $\Gamma$ is a function of two momenta while $D$ is a function of a single momentum,  the
equation for 
$\Gamma$ is  somewhat simpler than the gap equation for $D$. 
This is because the
equation for $\Pi$ is a self-consistent equation: $\Pi$ enters the 
calculation of the dressed propagator which
in turns is used to calculate $\Pi$. In contrast, the 4-point 
function $\Gamma$ does not enter the determination
of the kernel $\Lambda$ of the Bethe-Salpeter  equation.

For the construction of the mass counterterm, it will be later convenient
to consider also the derivative of the self-energy with respect to 
the mass. In
perturbation theory, the mass enters only the free propagator $D_0$, and
 $\del D_0/\del m^2= -D_0^2$. Thus the diagrams
for $m^2\del \Pi/\del m^2$ are those obtained by inserting $m^2$ in one of the internal
propagators. A simple analysis then shows that
\beq\label{delPix}
m^2\frac{\del \Pi(P)}{\del m^2}=-\frac{m^2}{2}\int_K \Gamma(P,K) D^2(K).
\eeq
This equation can be understood as follows: Consider first those diagrams contributing
to $m^2(\del \Pi/\del m^2)$ which contain no self--energy insertions (skeletons). 
It is then not hard to see that the contribution
of these diagrams is given by the r.h.s. of Eq.~(\ref{delPix}) in which $D$ is replaced 
by $D_0$, and similarly $\Gamma$ is replaced by $\Gamma_0$. Eq.~(\ref{delPix}) then follows by noticing that
the complete set of diagrams contributing to $m^2(\del \Pi/\del m^2)$ is
 obtained by replacing
in the skeleton diagrams identified above the free propagators $D_0$ by the full ones $D$.

Eq.~(\ref{delPix}) may be also obtained by the following algebraic
manipulations that will be used later on. Using Eq.~(\ref{Lambdadef}) and the fact that 
$\del D/\del m^2=-D^2(1+\del
\Pi/\del m^2)$, we get
\beq\label{m2dPidm2bx}
m^2\frac{\del \Pi(K)}{\del m^2}= -\frac{m^2}{2}\int_Q \Lambda(K,Q)D^2(Q)
\left( 1+ \frac{\del \Pi(Q)}{\del m^2} \right),
\eeq
which we can rearrange  as follows
\beq\label{m2dPidm2cx}
\int_Q\left[\delta(K-Q)+\frac{1}{2}\Lambda(K,Q)D^2(Q)\right]\frac{\del \Pi(Q)}{\del
m^2}=-\frac{1}{2}\int_Q \Lambda(K,Q)D^2(Q).
\eeq
Now, the Bethe-Salpeter equation (\ref{BSdef}) can be written as 
\beq\label{G0cx}
\int_Q\left[\delta(P-Q)+\frac{1}{2}\Lambda(P,Q)D^2(Q)\right]\Gamma(Q,K)=\Lambda(P,K),
\eeq
or equivalently as
\beq\label{G0cy}
\int_K\left[\delta(P-K)-\frac{1}{2}\Gamma(P,K)D^2(K)\right]\Lambda(K,Q)=\Gamma(P,Q).
\eeq
From these, one easily obtains 
\beq
\int_K\left[\delta(P\!-\!K)-\frac{1}{2}\Gamma(P,K)D^2(K)\right]
\left[\delta(K\!-\!Q)+\frac{1}{2}\Lambda(K,Q)D^2(Q)\right] =\delta(P\!-\!Q).
\eeq
Using these equations, one easily transforms Eq.~(\ref{m2dPidm2cx}) into Eq.~(\ref{delPix}). 
Renormalization counterterms will slightly modify the form of this equation, as we shall see
shortly.

\subsection{Renormalization: generalities}

In $\phi^4$ theory in four dimensions, we know from power counting 
that, aside from the thermodynamical potential itself,  only the 2-point
    and the 4-point functions can be divergent when computed in perturbation
theory. 
Thus, one expects the subdivergences in any diagram of perturbation theory to be of two
types, associated respectively with divergences of a 2-point and a 4-point function. An important
class of  4-point subdivergences in a self-energy diagram will contain as external lines those of
the self--energy itself and those obtained by cutting a line (associated with a free propagator)
 which does not belong to a cycle of lines. The  analysis done previously indicates that
this is nothing but a contribution to $\Gamma$. One should therefore expect a strong interplay
between the vertex subdivergences of the gap equation and those of the Bethe-Salpeter equation.

The 
divergences of the $n$-point functions can be absorbed in local counterterms corresponding to  a redefinition of the
parameters of the  lagrangian. The initial theory depends on three parameters, the bare mass and coupling constant,
$m_B$ and
$\lambda_B$ respectively, and an ultraviolet cut-off. The  relations  between
the renormalized and the bare parameters (denoted by a subscript
$B$) are the usual ones:
\beq
\phi_B=\sqrt{Z}\phi,\quad
Zm_B^2=m^2+\delta m^2,\quad Z^2 \lambda_B=Z_\lambda \lambda=\lambda+\delta \lambda,\quad \delta Z=Z-1.
\eeq
The parameters of the renormalized theory are the renormalized mass and coupling constant, $m$ and $\lambda$
respectively, and the counterterms $\delta Z$, $\delta m^2$, $\delta \lambda$ are supposed to be functions of $m$,
$\lambda$ and the cut-off. They are such that physical observables, expressed in terms of renormalized quantities, have a
finite limit when the cut-off goes to infinity. 

The relation between bare and renormalized propagators reads:
\begin{eqnarray}
D_B = ZD
\end{eqnarray}
and we define bare and renormalized self-energies by
\beq\label{PIPiBdef}
D^{-1}_B(K)=K^2+m_B^2+\Pi_B\qquad D^{-1}(K)=K^2+m^2+\Pi.
\eeq
Thus $\Pi$ and $\Pi_B$ are related by
\beq
\Pi=Z\Pi_B+K^2\delta Z+\delta m^2.
\eeq
For the skeleton diagrams, we have the identity
\beq\label{renormPhi}
\Phi[\lambda_B,D_B]=\Phi[Z_\lambda \lambda,D].
\eeq
This follows from the fact that to each vertex of $\Phi$ are attached four propagators. Thus, in
$\Phi$,  one can replace  $D_B=ZD$ by $D$ provided   one  replaces at the same time $\lambda_B$ by
$Z^2\lambda_B=Z_\lambda \lambda$.   This relation (\ref{renormPhi}), together with the definitions
above, allows us to  express the thermodynamical potential 
$\Omega$ in terms of renormalized quantities
\beq\label{thermopotential}
\beta(\Omega[D]-\Delta\Omega)=\2\int_K\log D^{-1}(K)-\2\int_K\left(  \Pi(K) -\delta Z  K^2 - \delta
m^2 \right) D(K) +\Phi[Z_\lambda \lambda,D],\nn
\eeq
where $\Delta\Omega$ is an infinite constant,  independent of the temperature, whose role is to absorb the
global divergence which remains in $\Omega$ once the 2-point and 4-point function subdivergences have been properly
eliminated.  When expressed  in terms of bare quantities, the thermodynamical  potential is stationary with respect to
variations of
$D_B$, at fixed bare parameters (see Eq.~(\ref{selfcons})). One can verify that a similar property holds
when
$\Omega$ is expressed in terms of the renormalized propagator, i.e., the expression
(\ref{thermopotential}) is stationary with respect to variations of $D$ at fixed renormalized parameters
(and therefore fixed values of the counterterms). We have also
$\d \Phi/\d D_B = (1/2)\Pi_B$, and therefore:
\beq
\frac{\d \Phi[Z_\lambda \lambda,D]}{\d D}=\2Z\Pi_B,
\eeq
with $\Pi_B$ evaluated with the propagator $D$. Given the relation (\ref{PIPiBdef}) between $\Pi_B$ and $\Pi$, 
one then obtains the  gap equation  in the form:
\begin{equation}\label{eq:gap}
\Pi(K)=\delta Z K^2+\delta m^2+2\frac{\d \Phi[Z_\lambda \lambda,D]}{\d D}.
\end{equation}

Consider finally the Bethe-Salpeter equation. In terms of bare quantities it reads:
\beq
\Gamma_B(P,K)=\Lambda_B(P,K)-\2\int_Q \Lambda_B(P,Q) D_B^2(Q)\Gamma_B(Q,K), 
\eeq
where 
\beq
\Lambda_B(P,K)=4\frac{\d^2\Phi[\lambda_B,D_B]}{\d D_B(P)\d D_B(K)}.
\eeq
In terms of renormalized quantities, this is simply
\beq\label{RenormBS}
\Gamma(P,K)=\Lambda(P,K)-\2\int_Q \Lambda(P,Q) D^2(Q)\Gamma(Q,K), 
\eeq
with
\beq\label{Lambda0}
\Lambda(P,K)=Z^2\Lambda_B(P,K)=4\frac{\d^2\Phi[Z_\lambda \lambda,D]}{\d D(P)\d D(K)}
\eeq
Note that 
$\Lambda$ is not a finite quantity: as we shall see,
it contains coupling constant counterterms
which are needed to remove 
the divergences in the integral equation for $\Gamma$. 

Related to the Bethe-Salpeter equation is the equation for $\del\Pi/\del m^2$. In terms of bare
quantities this is Eq.~(\ref{delPix}). For the renormalized theory we  have:
\beq\label{m2dPidm2bR}
m^2\frac{\del \Pi(K)}{\del m^2}= -\frac{m^2}{2}\int_Q \Lambda(K,Q)D^2(Q)
\left( 1+ \frac{\del \Pi(Q)}{\del m^2} \right)+\delta m^2 ,
\eeq
where we have used the fact that $\delta m^2/m^2$ can be chosen to be a constant independent
of $m^2$. 
This equation will be used to determine the mass counterterm from the requirement that $\del\Pi/\del m^2$ be finite. 

Let us now summarize the procedure that we shall follow in order to identify properly the
subdivergences.  We shall expand the self-energy in powers of the coupling constant, keeping at a
given order only those  diagrams which are  generated by iterating the gap
equation. 
Then we shall perform a standard analysis of the subdivergences and show that they can be
eliminated order by order by appropriate local counterterms. 
We shall be helped in this analysis by the consideration of the Bethe-Salpeter
equation which will lead to the  determination of the coupling constant counterterm
$\delta\lambda$, and the equation for
$\del
\Pi/\del m^2$ which will fix the mass counterterm
$\delta m^2$.  The last counterterm,
$\delta Z$, associated with field normalization,  will be obtained independently by considering an
equation for the asymptotic part of the propagator.
Once the counterterms are known, one can verify that the gap equation is finite,
and we shall present a well defined procedure to obtain such an equation.
One can also verify that the temperature dependent infinities go away: these appear typically as products of finite temperature
contributions, which are finite, by zero temperature subdiagrams, which are infinite, but are renormalized by zero
temperature counterterms. 
In summary, once the subdivergences have been properly eliminated, all apparent 
temperature dependent divergences are also eliminated.

\section{The two-loop approximation\label{sec:twoloop}}
\setcounter{equation}{0}

In this section, we discuss the approximation obtained by keeping in $\Phi$ only the two-loop skeleton. The presentation is
more detailed than needed for the solution of the problem, which is well known
\cite{Dolan:qd,Drummond:1997cw}.
However, many features of the renormalization
of a self-consistent approximation already appear 
in this simple example and can be made explicit with analytical
calculations. This two-loop  example constitutes therefore an excellent preparation for the generalizations to be
done in the next sections.

The contribution of the two-loop skeleton is
\beq\label{Phitadpole}
\Phi\,=\frac{\lambda_B}{8}\left(\int_P D_B(P)\right)^2=
\frac{\lambda+\delta \lambda}{8}\left(\int_P D(P)\right)^2.
\eeq
The self-energy,  obtained from Eq.~(\ref{eq:gap}), reads then:
\beq\label{Pitad}
\Pi = \frac{\lambda+\delta\lambda}{2}\int_P D(P) +\delta m^2
\eeq
where $D^{-1}(P)=P^2+m^2+\Pi $, and $\Pi$ is in this case a constant (so that
$\delta Z=0$ and $D_B=D$). 
Finally the  kernel of the Bethe-Salpeter equation is
\beq
\Lambda=\lambda_B=\lambda+\delta \lambda,
\eeq
and the Bethe-Salpeter  equation itself reads:
\beq\label{BStad}
\Gamma=\Lambda-\frac{\Lambda}{2}\,\Gamma\int_P
D^2(P),
\eeq
with $\Gamma$ a constant. This equation can be conveniently rewritten in the form:
\beq\label{BStadbis} 
\frac{1}{\Gamma}=\frac{1}{\Lambda}+\frac{1}{2}\int_P
D^2(P).
\eeq

The integrals above are ultraviolet divergent. We define them through dimensional regularisation, with
$d=4-2\epsilon$  and $\mu$ the renormalization scale. The dimension of the 4-point functions
$\Lambda$
 and $\Gamma$
 becomes then  that of the bare coupling constant $\lambda_B$ in $d$ dimensions, i.e.,  $\mu^{2\epsilon}$. It is convenient to
rescale these functions so as to keep them dimensionless:
\beq
\Lambda\longrightarrow \mu^{2\epsilon}\Lambda,\quad \Gamma\longrightarrow \mu^{2\epsilon}\Gamma,\quad
\lambda_B\longrightarrow \mu^{2\epsilon} (\lambda+\delta \lambda).
\eeq
The Bethe-Salpeter  equation reads then as in Eq.~(\ref{BStad}), with a factor $\mu^{2\epsilon}$
multiplying  the integral.
As for the gap equation it becomes
\beq\label{Pitad1}
\Pi = \frac{\Lambda}{2}\mu^{2\epsilon}\int_P D(P) +\delta m^2, 
\eeq
and $\Pi$ has the same dimension as $m^2$. 

The  foregoing analysis will involve 
Feynman diagrams calculated with the free propagator: $D_0(P)=(P^2+m^2)^{-1}$. 
The needed integrals are (with  $\bar\mu^2=4\pi {\rm
e}^{-\gamma_E}\mu^2$):
\beq\label{int10}
{\cal B}_0(m^2)\equiv\mu^{2\epsilon} \int_P D_0(P)=-\frac{m^2}{16\pi^2} \left\{ 
\frac{1}{\epsilon}
-\ln\frac{m^2}{\bar\mu^2}+ 1\right\}\equiv \frac{a_0}{\epsilon}+b_0,
\eeq
and 
\beq\label{int20}
{\cal B}_1(m^2)\equiv\mu^{2\epsilon}
\int_P D^2_{0}(P)\,=\,
\frac{1}{16\pi^2} \left\{ \frac{1}{\epsilon}
-\ln\frac{m^2}{\bar\mu^2}\right\}=-\frac{\del {\cal B}_0}{\del m^2}\equiv \frac{a_1}{\epsilon}+b_1.
\eeq
Note that ${\cal B}_1$ is dimensionless and that its divergent piece $a_1/\epsilon$ is 
independent of $m$ ($a_1=1/16\pi^2$). 

At finite temperature, these two integrals receive additional finite contributions, respectively $b_0^T$ and $b_1^T$, with
\beq
b_0^T=\int \frac{{\rm d}^3p}{(2\pi)^3}\frac{n(\epsilon_p)}{\epsilon_p},
\eeq
and $b_1^T=-\del b_0^T/\del m^2$.

Before we proceed, let us comment on a typical difficulty that we are faced with. Since the
divergence in Eq.~(\ref{Pitad}) is a local one, one could, naively, attempt to absorb it in the mass
counterterm $\delta m^2$. However, if one does this, one obtains a counterterm whose
divergent part depends on the self-energy $\Pi$. This is not desirable since, at finite temperature,
$\Pi$ depends on the temperature, and that would  introduce temperature dependent counterterms in the calculation. 
Besides, one expects on general grounds to be a
ble to put the mass counterterm in the
form $\delta m^2=m^2 C(\lambda,\epsilon)$, with $C$ independent of $m$ \cite{Collins:xc}.
Clearly, the
$\Pi$--dependent mass counterterm just defined would not have this property, 
as easily seen by taking a derivative with respect to $m^2$.
In fact  the infinities in Eq.~(\ref{Pitad})
result from subdivergences which are best exhibited by iterating the  gap equation. And to properly identify
$\delta\lambda$ and $\delta m^2$, one needs to carefully disentangle these various subdivergences.

We are therfore led to  analyze the formal solutions of the gap equation (\ref{Pitad1}), and of the
Bethe-Salpeter equation (\ref{BStad}), that are obtained by the successive iterations of these equations. Such
iterations construct the solutions as powers series in the coupling constant $\lambda$. These series  coincide
with those one would obtain by applying the rules of  ordinary perturbation theory, and restricting 
the  Feynman diagrams to belong to a certain topology. The solutions after
$n$ iterations, to be denoted respectively by $\Pi^{(n)}$ and $\Gamma^{(n)}$, are of order $\lambda^{n+1}$,
 with each iteration bringing in a
new power of $\lambda$, and associated new divergences.   The renormalization
consists in showing that, given the mass and coupling constant counterterms
determined at the previous iterations, of order $\lambda^n$ at most, the new
divergences can be absorbed in a new contribution to these counterterms, of
order $\lambda^{n+1}$. Note that whether this can be done or not is not guaranteed a priori by the
standard proof of perturbative renormalizability,  since at any order $\lambda^n$ only a subset of terms
contributing at this order is included.

\subsection{Iterative
 solution of the Bethe-Salpeter equation \label{sec:BStadpole}}

We consider first the Bethe-Salpeter equation (\ref{BStad}) in which we replace the propagator $D$ by   the
free propagator $D_0$. We call $\Gamma_0$ the corresponding 4-point function, that is:
\beq\label{BStad0}
\Gamma_0=\Lambda-\frac{\Lambda}{2}\,\Gamma_0\mu^{2\epsilon}\int_P
D^2_0(P)=\Lambda-\frac{\Lambda}{2}\Gamma_0{\cal B}_1(m^2).
\eeq
 As we have just mentioned, we shall construct the counterterms by building the solution  of
Eq.~(\ref{BStad0}) as a formal series in powers of the coupling constant $\lambda$. This is
obtained by iteration, keeping at each iteration $k$   only terms which contribute to $\Gamma_0$
up to order
$\lambda^{k+1}$. Thus, 
$\Gamma^{(n)}_0$,  the value of $\Gamma_0$ obtained after $n$ iterations,  is of order $\lambda^{n+1}$,
and so is the counterterm $\delta\lambda_n$ that needs to be adjusted at iteration $n$.
Clearly, we can write
\beq\label{Lambdacounterterm}
[\Lambda]_{[n+1]}=\lambda+\sum_{k=1}^n\delta\lambda_k.
\eeq
where $\delta\lambda_k$ is the counterterm 
of order $\lambda^{k+1}$. The notation $[\cdots]_{[n]}$, to be used throughout, 
 indicates that only the terms up
 to order $\lambda^n$ are to be kept within the brakets.

\begin{figure}[htb]
\begin{center}
\includegraphics[width=4cm]{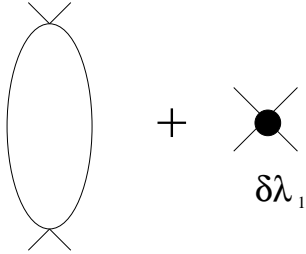}
\caption{The counterterm $\delta \lambda_1$ eliminates the divergence of the one-loop contribution to the 4-point function
$\Gamma_0^{(1)}$.\label{fig:1loop_ren}}
\end{center}
\end{figure}

 In leading order, the
solution $\Gamma^{(0)}_0$ is simply the renormalized coupling constant
\beq
\Gamma^{(0)}_0=\lambda.
\eeq
The first iteration gives
\beq\label{Gamma01}
\Gamma^{(1)}_0=\lambda+\delta \lambda_1-\lambda^2\frac{{\cal B}_1}{2}=\lambda+\delta \lambda_1-\frac{\lambda^2}{2}\left(\frac{a_1}{\epsilon}+b_1\right).
\eeq
As anticipated, there is a divergent contribution given by the one-loop diagram in Fig.~\ref{fig:1loop_ren}.
 This divergence is absorbed in the counterterm 
 $\delta \lambda_1$. Using minimal subtraction, one gets:
\beq
\delta \lambda_1=\frac{\lambda^2 a_1}{2\epsilon}.
\eeq
Note that we have ignored in (\ref{Gamma01}) 
the terms $\sim \lambda \delta \lambda_1$ and $\sim \delta \lambda_1^2$ which
are of order higher than $\lambda^2$. Note also that $\Gamma^{(1)}_0$ is only a
part of the 4-point function that would be obtained in perturbation theory at
order $\lambda^2$. This is so because, as we have already mentioned,  the Bethe-Salpeter equation performs
a resummation  in only one out of three  possible channels. The contributions of other channels are recovered only when 
included in the kernel $\Lambda$ (see next section).

\begin{figure}[htb]
\begin{center}
\includegraphics[width=8cm]{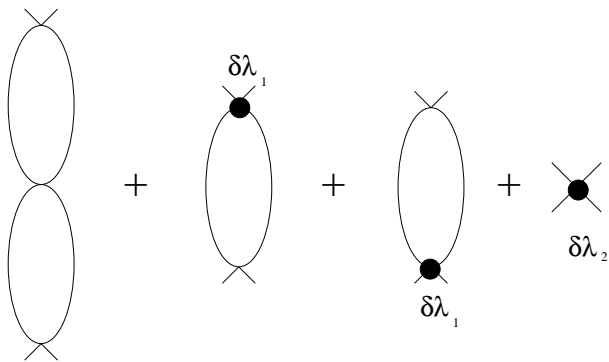}
\caption{The counterterm $\delta \lambda_2$ eliminates the
global divergence which remains in the two-loop contribution to  $\Gamma_0^{(2)}$ (diagram on left) after the subdivergences
have been removed by the  two middle diagrams involving $\delta\lambda_1$.
\label{fig:2loop_ren}}
\end{center}
\end{figure}

Doing the second iteration and keeping only terms up to order $\lambda^3$, one obtains
\beq
\Gamma^{(2)}_0&=& \left[\Lambda-\Lambda\frac{{\cal B}_1}{2}\Gamma^{(1)}_0\right]_{\left[ 3\right]}\nonumber\\
&=&\lambda+\delta \lambda_1+\delta \lambda_2-\lambda\left(\lambda+2\delta
\lambda_1\right)\frac{{\cal B}_1}{2}+ \lambda^3\frac{{\cal B}_1^2}{4},
\eeq
where the notation $[\cdots]_{[n]}$ is that of Eq.~(\ref{Lambdacounterterm}). 
Note that although there is a single counterterm to adjust, namely $\delta \lambda_2$, one
needs to keep combinations of terms which where ignored at the previous
iteration, such as $\lambda \delta \lambda_1$: such terms are needed to
 remove the subdivergences that are now present, as can be seen in the diagrams displayed in Fig.~\ref{fig:2loop_ren}.
Observe that in the cancellation of subdivergences, the term involving the product of $\delta\lambda_1$ and the finite part
of
${\cal B}_1$ also cancels. 
The global divergence that remains, and  that is absorbed in $\delta\lambda_2$, depends therefore only on $a_1$. A
simple calculation gives
\beq
\delta\lambda_2=\lambda\left(\frac{\lambda a_1}{2\epsilon}\right)^2.
\eeq

\begin{figure}[htb]
\begin{center}
\includegraphics[width=6cm]{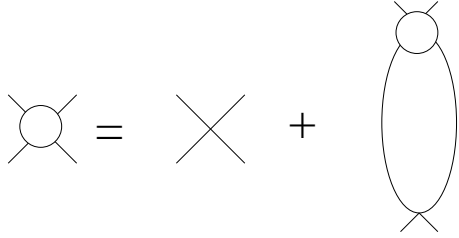}
\caption{Diagrammatic interpretation of the Bethe-Salpeter equation\label{fig:BS}.}
\end{center}
\end{figure}
\begin{figure}[htb]
\epsfysize=7.cm
\centerline{\epsffile{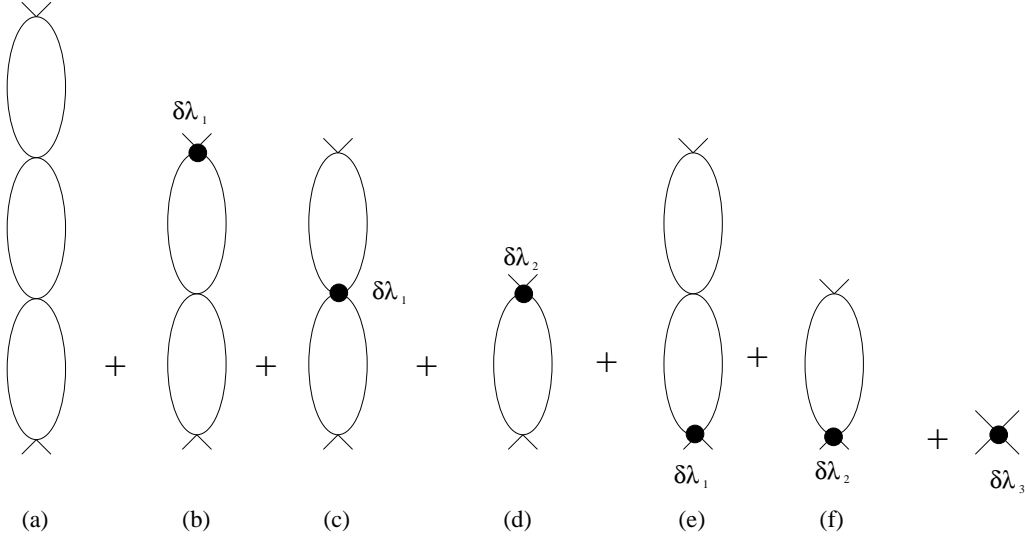}}
\vspace{3mm}
\caption[a]{Contributions to $\Gamma^{(2)}_0$ at order $\lambda^3$, illustrating the elimination of 
 the various subdivergences of the diagram (a) by counterterms determined at the first two iterations 
(i.e., $\delta\lambda_1$ and $\delta\lambda_2$); these leave a global divergence absorbed in
$\delta\lambda_3$. Diagrams (b,c,d) represent counterterms for subdivergences which do
not involve the lower vertex of (a); diagrams (e) and (f) eliminate the  subdivergences
involving the lower vertex of (a).\label{bs_it}}
\end{figure}

Quite generally, at iteration  $n$ one gets
\beq\label{BSitern}
\Gamma^{(n)}_0=\left[\Lambda -\Lambda  \Gamma^{(n-1)}_0\frac{{\cal B}_1}{2}\right]_{[n+1]}.
\eeq
The term $\Gamma^{(n-1)}_0$ is assumed to be finite. It contains all the counterterms
which eliminate the subdivergences  that do not involve the lowest vertex in  Fig.~\ref{fig:BS}. The
factor $\Lambda$ multiplying 
 it brings the counterterms needed to remove the subdivergences involving the lower vertex.
After all the subdivergences have been removed by the counterterms determined 
in the $(n-1)$ previous iterations, there remains a global divergence which is cancelled by the adjustment of the
contribution $\delta\lambda_n$ to   $\Lambda$ (which affects only the first $\Lambda$ in the r.h.s. of Eq.~(\ref{BSitern})).
An illustration at order
$\lambda^3$ is provided in Fig.~\ref{bs_it}. 

One can rewrite Eq.~(\ref{BSitern}) as follows:
\beq\label{iterBSn}
\Gamma^{(n)}_0=\left[\frac{\Lambda}{1+\Lambda\frac{{\cal B}_1}{2}}\right]_{[n+1]}
=\left[\Gamma_0\right]_{[n+1]}.
\eeq
The last equality emphasizes the
fact that
$\Gamma^{(n)}_0$ is nothing but the perturbative solution of order
$\lambda^{n+1}$, restricted to the particular resummation of the chain of bubbles in the $t$-channel.
In order to determine the general form of the counterterms, it is in fact more convenient to invert this relation:
\beq\label{iterLambdaBSn}
\left[\Lambda\right]_{[n+1]}=\left[\frac{\Gamma_0}{1-\Gamma_0\frac{{\cal
B}_1}{2}}\right]_{[n+1]},
\eeq
and exploit the fact that $\Gamma_0$ is finite. 
Then, it is not hard to get
\beq\label{deltalambdak}
\lambda=\frac{\Gamma_0}{1-\frac{b_1}{2}\Gamma_0}\,\, ,\qquad \delta
\lambda_n=\lambda\left(\frac{\lambda a_1}{2\epsilon}\right)^n,
\eeq
where $a_1$ and $b_1$ are defined in Eq.~(\ref{int20}).

At this point we note that the counterterms depend only on $a_1$, that is, they are independent of the
mass. It follows that the Bethe-Salpeter equation (\ref{BStad0}) written with the full
propagator 
$D$ instead of the free propagator $D_0$, can be made finite with the same counterterms as
those determined for $\Gamma_0$. This may also be seen by observing that any insertion of a
self-energy in the propagators of the Bethe-Salpeter equation alters only the finite part of the
integral in Eq.~(\ref{BStad}).

Let then $\Gamma$ be the  solution obtained
with the full propagator
$D$. From the first of Eqs.~(\ref{deltalambdak}), and a similar equation for $\Gamma$ (obtained by replacing simply $m^2$ by $m^2+\Pi$), one easily gets
\beq\label{BStadter} 
\frac{1}{\lambda}=\frac{1}{\Gamma_0}-\2b_1(m^2)=\frac{1}{\Gamma}-\2b_1(m^2+\Pi).
\eeq
Thus, $\Gamma$ and $\Gamma_0$ differ indeed only by a finite part:
\beq
\frac{1}{\Gamma}-\frac{1}{\Gamma_0}=\frac{1}{32\pi^2}\ln\frac{m^2}{m^2+\Pi},
\eeq
where we have used the explicit form of $b_1(m^2)$ given in Eq.~(\ref{int20}).

\begin{figure}[htb]
\epsfysize=4.cm
\vspace{1cm}
\includegraphics[width=4cm]{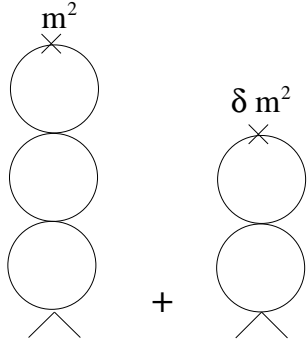}
\caption[a]{Diagrammatic representation of the two terms in the right hand side of
Eq.~(\ref{m2dPidm2a0}). \label{fig:dPidm2}}
\end{figure}

\subsection{The equation for the mass counterterm}

We now turn to the equation satisfied
 by $\del \Pi/\del m^2$, whose structure, as discussed in Sect. II,
  is very close to that of the Bethe-Salpeter equation. 
This  will allow  us to determine the counterterm $\delta m^2$,
 which we assume of the form
$\delta m^2=m^2 C(\lambda,\epsilon)$. From Eq.~(\ref{m2dPidm2bx}) (or directly from Eq.~(\ref{Pitad})), one easily gets:
\beq\label{m2dPidm2a0}
m^2\frac{\del \Pi}{\del m^2}&=& -\left(1+\frac{\Lambda}{2}{\cal
B}_1\right)^{-1}\,\frac{\Lambda}{2}{\cal B}_1 m^2+\left(1+\frac{\Lambda}{2}{\cal
B}_1\right)^{-1}\,\delta m^2,
\eeq
where ${\cal B}_1$ stands here for ${\cal B}_1(m^2+\Pi)$. 
This equation has a simple diagrammatic interpretation: each term represents a tower of bubbles, with $m^2$ or
$\delta m^2$ inserted in the upper most bubble (see Fig.~\ref{fig:dPidm2}). The corresponding diagrams are
easily generated by expanding the right hand side of this equation
in powers of $\lambda$. 
A simple analysis then reveals that the subdivergences
associated with coupling constant counterterms cancel out; that is, the potential subdivergences are removed by
the same counterterms as introduced in solving the Bethe-Salpeter equation. Such subdivergences in the
towers of bubbles of Fig.~\ref{fig:dPidm2}  are those which  do not involve the upper loop with  $m^2$ or
$\delta m^2$ inserted. The subdivergences involving the upper loops are remove by adjusting the mass counterterm
$\delta m^2$ at each iteration. The elimination of coupling constant subdivergences can be made explicit
by using the Bethe-Salpeter equation (\ref{iterBSn})  to write:
\beq\label{m2dPidm2a1}
m^2\frac{\del \Pi}{\del m^2}= -\frac{\Gamma}{2}{\cal B}_1 m^2+\left(1-\frac{\Gamma}{2}{\cal
B}_1\right)\,\delta m^2,
\eeq
where we have used the identity
\beq
\left(1+\frac{\Lambda}{2}{\cal
B}_1\right)^{-1} \, \left(1-\frac{\Gamma}{2}{\cal
B}_1\right)=1.
\eeq
The mass counterterms can then be easily determined by expanding the r.h.s. of  Eq.~(\ref{m2dPidm2a1}) in powers of $\lambda$, identifying at each
order ³ the divergent part of
$\delta m^2$. In leading order we get
\beq\label{deltam0}
\delta m^2_0= m^2 \frac{\lambda a_1}{2\epsilon}.
\eeq
At next order, we have:
\beq\label{dPidm2it1}
m^2\frac{\del \Pi^{(1)}}{\del m^2}=-m^2 \frac{\Gamma^{(1)}}{2}{\cal B}_1+\delta
m^2_0
\left(1-\frac{\lambda}{2}{\cal B}_1\right)+\delta m^2_1,
\eeq
where $\Pi^{(1)}$ and $\Gamma^{(1)}$ are the values of $\Pi$ and $\Gamma$ obtained at the first iteration of, respectively, the gap equation and the Bethe-Salpeter equation. Keeping only the divergent parts of order $\lambda^2$ one gets
\beq\label{deltam1surm2}
\frac{\delta m^2_1}{m^2}&=&\left[\left(\delta\lambda_1-\frac{\lambda^2}{2} {\cal
B}_1\right)\frac{{\cal B}_1}{2}\right]_{div} +\lambda \frac{\delta m^2_0}{m^2}\frac{{\cal
B}_1}{2}\nonumber\\
 &=&\left( -\frac{\lambda^2}{2} b_1\right)\frac{a_1}{2\epsilon}+ \frac{\delta
m^2_0}{m^2}\left(\frac{\lambda a_1}{2\epsilon}+\frac{\lambda b_1}{2}\right),
\eeq
where we have used Eq.~(\ref{Gamma01}) for $\Gamma^{(1)}$.
Observe the cancellation of the term involving the leading order mass counterterm $\delta m^2_0$ multiplied by the finite
part $b_1$, leaving the following formula for $\delta m^2_1$:
\beq\label{deltam21}
\frac{\delta m^2_1}{m^2}&=&\frac{\delta m^2_0}{m^2}\left(\frac{\lambda a_1}{2\epsilon}\right)\nonumber\\
&=&\frac{\delta\lambda_1}{2}\frac{a_1}{\epsilon}.
\eeq
In the second line we have written the formula which emerges naturally if one cancels 
the subdivergences in a different order (making use of the explicit expression (\ref{deltam0}) of $\delta m_0$).
This writing generalizes  to high orders. This can be seen by  rewriting Eq.~(\ref{m2dPidm2a1}) as
follows:
\beq\label{delPtad}
m^2\frac{\del \Pi}{\del m^2}= -m^2\frac{\Gamma}{2}{\cal B}_1+\frac{\Gamma}{\Lambda}\delta m^2 .
\eeq
Clearly, the divergence in the right hand side is cancelled if we choose 
\beq\label{deltam2Lambda}
\frac{\delta m^2}{m^2}=\frac{\Lambda}{2} \frac{a_1}{\epsilon}.
\eeq
We have obtained earlier $\Lambda$ as a power series in the coupling constant
(see Eqs.~(\ref{Lambdacounterterm}) and (\ref{deltalambdak})). Similarly we can write
\beq
[\delta m^2]_{[n+1]}=\sum_{k=0}^n\delta m^2_k,
\eeq 
with 
\beq\label{deltamka}
\delta m^2_k=m^2\left(\frac{\lambda a_1}{2\epsilon}\right)^{k+1}.
\eeq

As was the case for the Bethe-Salpeter equation, it does not matter for the determination of the
divergent part of the mass counterterm whether the free propagator $D_0$, or the full propagator $D$ is used
in  Eq.~(\ref{delPtad}): this is a priori clear from the diagrammatic structure in Fig.~\ref{fig:dPidm2} and is obvious in
Eq.~(\ref{deltam2Lambda}). But Eqs.~(\ref{m2dPidm2a0}) or (\ref{delPtad})  for
$\del\Pi/\del m^2$,  including the finite parts, hold  only if the
full propagator is used (as is clear from their derivation).

\subsection{Iterative solution of the gap equation\label{sec:itsolBStadpole}}

We turn now to the gap equation (\ref{Pitad1}), and  construct iteratively its solution as a formal series in powers of 
the coupling constant $\lambda$. At a given order in $\lambda$, $\Pi$ contains two types of subdivergences: those associated with the renormalization of the coupling constant, and those associated with the renormalization of the mass. 

 We shall verify that  the counterterms 
$\delta\lambda_n$ determined in solving  the
Bethe-Salpeter equation eliminate the subdivergence of the 
first kind, while those of the second kind are eliminated by the mass counterterm  
$\delta m^2_n$ obtained at the end of the previous subsection. 
\begin{figure}[htb]
\begin{center}
\includegraphics[width=10cm]{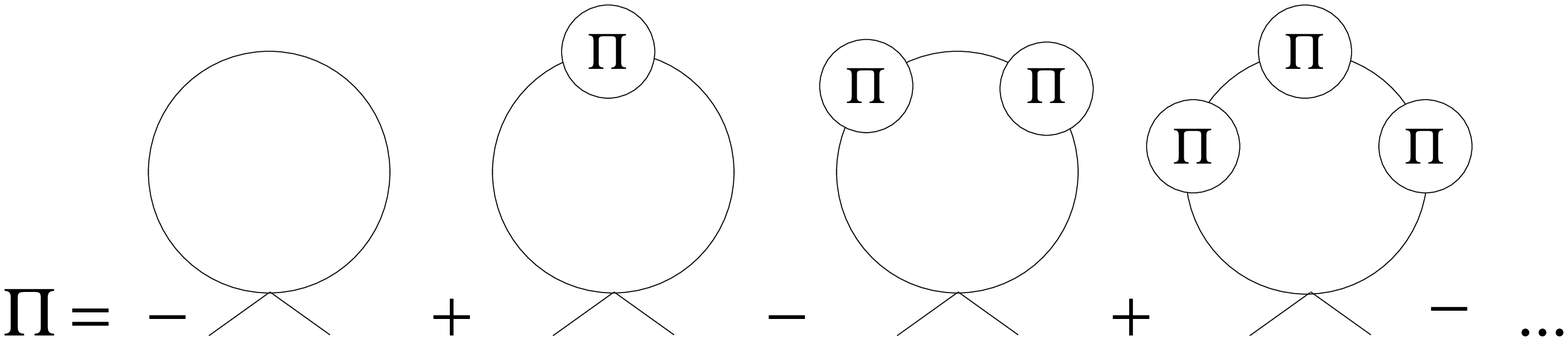}
\caption{Diagrammatic interpretation of the expansion (\ref{expansiongapeq})\label{fig:gap_dev}}
\end{center}
\end{figure}

We start by expanding the r.h.s. of the
gap equation in powers of $\Pi$:
\beq\label{expansiongapeq}
\Pi&=&\frac{\Lambda}{2}\int_P \frac{1}{P^2+m^2+\Pi}+\delta m^2\nonumber\\
 &=& \frac{\Lambda}{2}\sum_{k=0}(-1)^k{\cal B}_k \Pi^k+\delta m^2.
\eeq
The coefficients ${\cal B}_k $ are given by
\beq
{\cal B}_k=\mu^{2\epsilon}\int_P \frac{1}{(P^2+m^2)^{k+1}}=\frac{(-)^k}{k!}\frac{\del^k B_0}{\del
(m^2)^k}.
\eeq
The expansion (\ref{expansiongapeq}) is  represented diagrammatically in
Fig.\ref{fig:gap_dev}. Note that only ${\cal B}_0$ and ${\cal B}_1$ are ultraviolet divergent 
and, as we have seen in the previous subsection, in minimal subtraction  the counterterms $\delta\lambda$ and $\delta m^2$  
depend only on their divergent parts, respectively
$a_0$ and
$a_1$.  Each iteration brings in one power of $\lambda$ so that $\Pi^{(n)}$, the value of $\Pi$
at iteration
$n$, is of order $\lambda^{n+1}$. This means that at iteration $n\ge 1$ one can stop the expansion
(\ref{expansiongapeq}) at   $k=n$.

\begin{figure}[htb]
\begin{center}
\includegraphics[width=4cm]{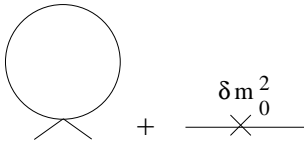}
\caption{The 1-loop contribution to $\Pi^{(0)}$ , and the corresponding  mass counterterm. \label{fig:1loop}}
\end{center}
\end{figure}

The
lowest order contribution,  $\Pi^{(0)}$,  is that of the 1-loop diagram in Fig.\ref{fig:1loop}:
\beq\label{Piiteration0}
\Pi^{(0)} = \frac{\lambda}{2}{\cal B}_0+\delta m^2_0
 = \frac{\lambda}{2}\left(\frac{a_0}{2}+b_0\right)+\delta m^2_0.
\eeq
There is no subdivergence in this diagram and the  global divergence 
is absorbed in the mass counterterm $\delta m^2_0$ given in Eq.~(\ref{deltam0})
(recall that $a_0=- m^2 a_1$).

\begin{figure}[htb]
\begin{center}
\includegraphics[width=10cm]{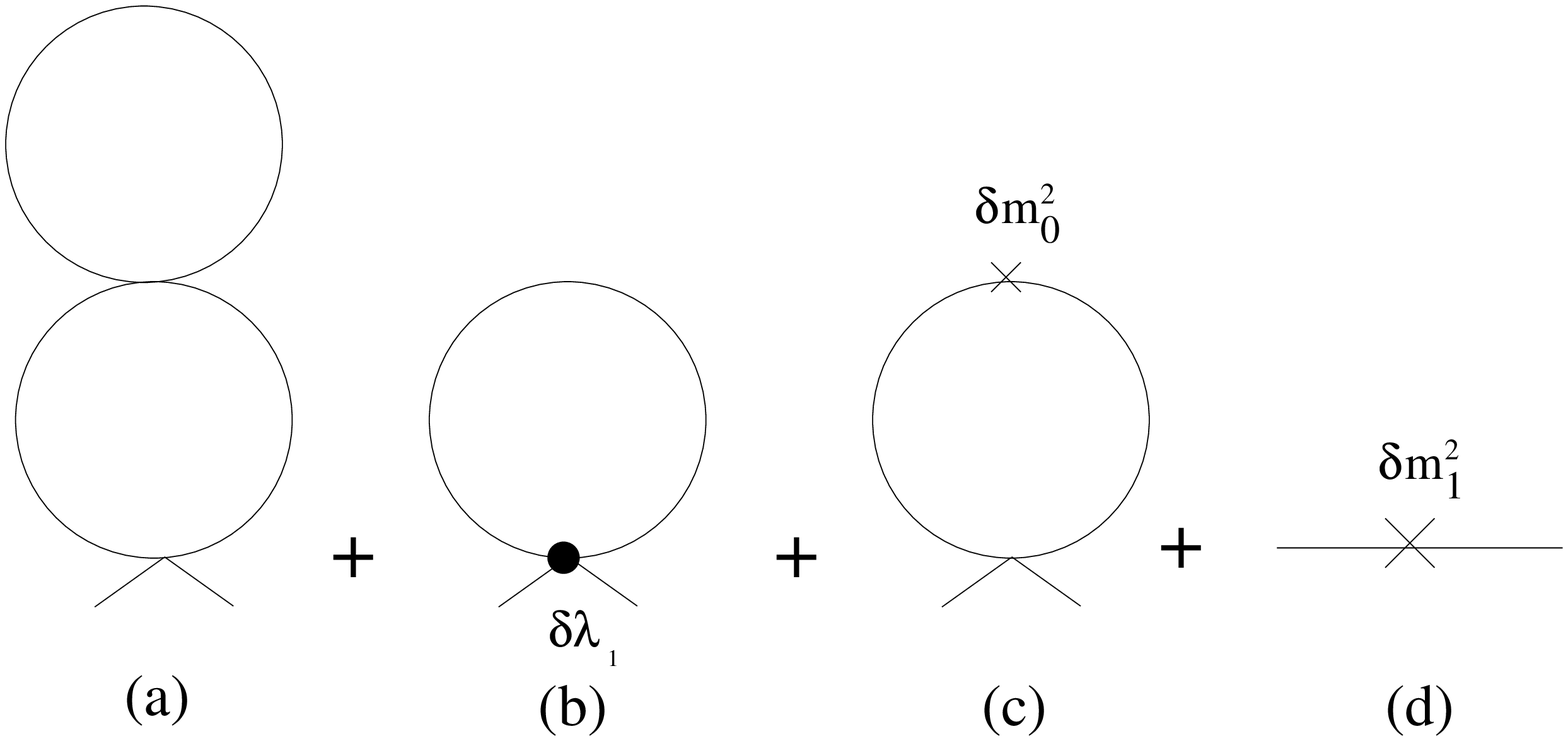}
\caption{The 2-loop contribution to $\Pi^{(1)}$, exhibiting two subdivergences.That associated 
with coupling constant renormalization (the lower loop) is eliminated by diagram (b) involving
$\delta\lambda_1$; that associated with mass renormalization (the upper loop in (a)) is eliminated by diagram
(c) involving $\delta m_0^2$. The remaining global divergence is absorbed in the mass counterterm $\delta
m_1^2$.
\label{fig:2loop}}
\end{center}
\end{figure}

The first iteration of the gap equation, in which we keep only the terms up to order $\lambda^2$, leads to 
\beq\label{deltam2}
\Pi^{(1)} &= &\left[- \frac{\Lambda}{2}{\cal B}_1\Pi^{(0)}\right]_{[2]}+\frac{1}{2}\left[ 
\Lambda\right]_{[2]}{\cal B}_0+\left[ 
\delta m^2\right]_{[2]}
\nonumber\\
&=& \frac{\lambda}{2}{\cal B}_0+\delta m^2_0
+\frac{{\cal B}_0}{2}\left(\delta \lambda_1-\frac{\lambda^2}{2}{\cal B}_1\right)-\delta
m^2_0\frac{\lambda}{2}{\cal B}_1+\delta m^2_1.
\eeq
In the second line we have grouped the terms so as 
to display the cancellation of divergences: one recognizes indeed in the last four terms, of order
$\lambda^2$,   the structure already met in Eq.~(\ref{dPidm2it1}) for $\del\Pi/\del m^2$ (recall that the
divergent part of
${\cal B}_0$ is the same as that of $m^2 {\cal B}_1$). This is illustrated in Fig.~\ref{fig:2loop}.

A further iteration will  allow us to uncover the generic features of the iteration procedure. Keeping only
terms up  to order
$\lambda^3$ on obtains:
\beq\label{Pi2emeit}
\Pi^{(2)}=\frac{\lambda}{2}{\cal B}_2\left(\Pi^{(0)}\right)^2-\frac{\lambda}{2} {\cal
B}_1\Pi^{(1)}-\frac{\delta\lambda_1}{2}{\cal
B}_1\Pi^{(0)}+\frac{1}{2}\left[\Lambda\right]_{[3]}{\cal B}_0+
\left[\delta m^2\right]_{[3]},
\eeq
\begin{figure}[htb]
\begin{center}
\includegraphics[width=14cm]{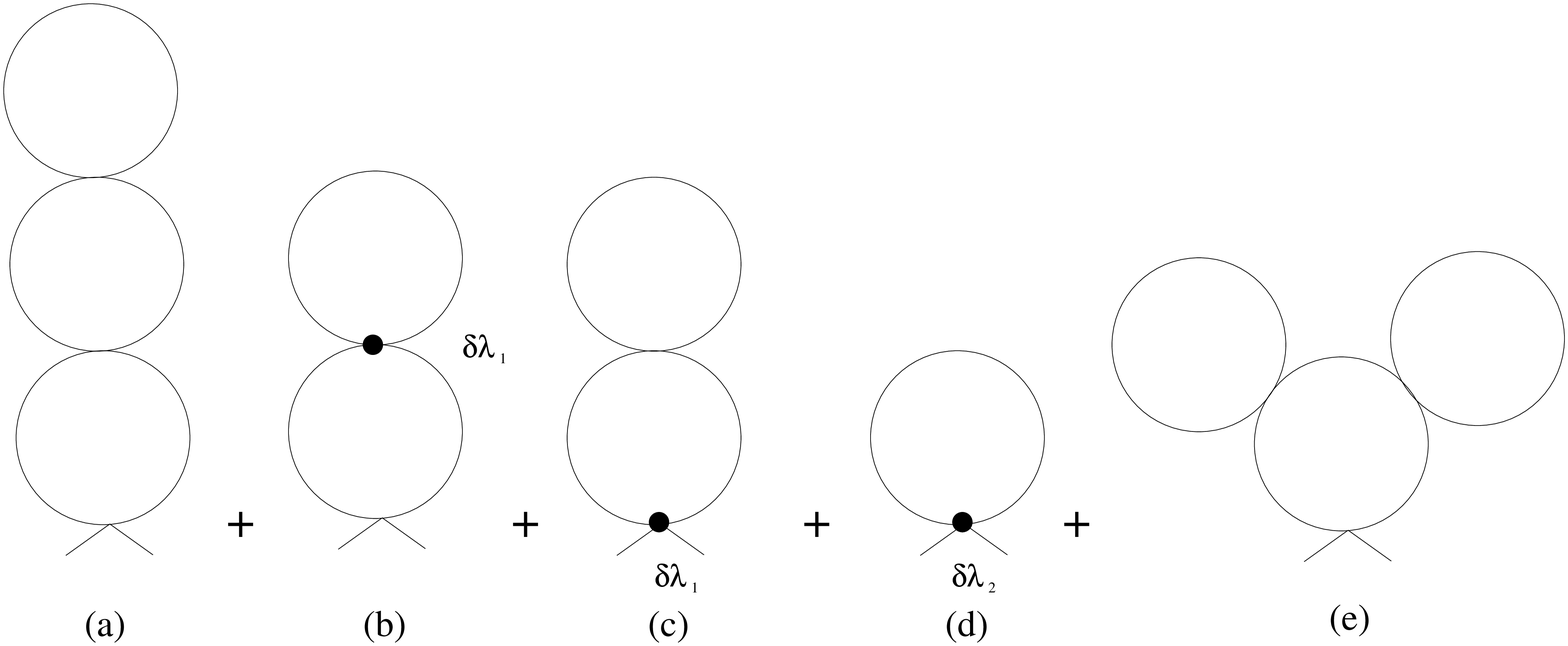}
\caption{The 3-loop contribution to $\Pi^{(2)}$, with its coupling constant counterterms (mass
counterterms are not drawn).\label{fig:3loop}}
\end{center}
\end{figure}
This iteration adds to $\Pi$ a 3-loop contribution, whose corresponding diagrams are displayed in
Fig.\ref{fig:3loop}. The first term in Eq.~(\ref{Pi2emeit}), quadratic in $\Pi$, is finite; all the
divergences are localized in the terms linear in $\Pi$ and to isolate them we need to express $\Pi^{(0)}$ and
$\Pi^{(1)}$ in terms of elementary integrals. Again, as was the case in the first iteration, we shall find that
the counterterms determined at the previous two iterations eliminate all the subdivergences but one, namely that
associated with diagram (a). This subdivergence is recognized as the global divergence of the order
$\lambda^3$ contribution to the 4-point function and it is absorbed in $\delta\lambda_2$. The remaining
divergence is a global one  and is eliminated by new mass counterterm $\delta m^2_2$. 

It is useful to observe that the new divergences which enter at each new iteration $n$, and which require the new
counterterms $\delta\lambda_n$ and $\delta m_n^2$, are entirely contained in the tower of bubbles such as diagram (a) in
Fig.~\ref{fig:3loop} . The other structures which develop as we proceed through the iterations, such as diagram (e) in
Fig.~\ref{fig:3loop},  are all made finite by previously determined counterterms (for diagram (e) this is due to the mass counterterm that is not drawn here). Then, it is not hard
to see that all the divergences in such a tower cancel in exactly the same way as they do in the equation for
$\del\Pi/\del m^2$. To make this explicit, note that we can rewrite Eq.~(\ref{Pi2emeit}) as follows
\beq\label{Piiteration2}
\Pi^{(2)}=\frac{\lambda}{2}{\cal
B}_2\left(\Pi^{(0)}\right)^2+\left[\left(1+\frac{\Lambda}{2}{\cal
B}_1\right)^{-1}\frac{\Lambda}{2}\right]_{[3]}{\cal B}_0+
\left[\left(1+\frac{\Lambda}{2}{\cal
B}_1\right)^{-1}\delta m^2\right]_{[3]}.
\eeq
The last two terms represent the contribution to $\Pi^{(2)}$ obtained by iterating only the term linear in
$\Pi$, thus generating the tower of bubbles such as that in Fig.~\ref{fig:3loop}(a), together with the
counterterms which make it finite. That the contribution of these last two terms of  Eq.~(\ref{Piiteration2}) is finite is made clear by comparing them to the right hand side of (\ref{m2dPidm2a0}),  observing that factor in front of ${\cal B}_0$ in Eq.~(\ref{Piiteration2}) is $\Gamma/2$ (see also Eq.~(\ref{m2dPidm2a1})), and that the divergent parts of ${\cal B}_0$ and $-m^2{\cal B}_1$ are the same.

One can summarize the iterative procedure by the following formula giving 
the solution $\Pi^{(n)}$ at iteration $n$ in terms of the (finite) solution $\Pi^{(n-1)}$ obtained at iteration $(n-1)$:
\beq
\Pi^{(n)}=\left[\frac{\Lambda}{2}\sum_{k=1}^n(-1)^k{\cal
B}_k\left(\Pi^{(n-1)}\right)^k\right]_{[n+1]}+\left[\Lambda\right]_{[n+1]}\frac{{\cal B}_0}{2}+
\left[\delta m^2\right]_{[n+1]}, 
\eeq
where we have separated the terms which contain $\Pi^{(n-1)}$ from those which contain only the  new
counterterms that need to be introduced at each iteration, and which are independent of $\Pi$.  It will
prove convenient to introduce a special notation for these two contributions. We set:
\beq\label{deftildePi0}
\tilde \Pi_0^{(n)}=\left[\frac{\Lambda}{2}\sum_{k=1}^n(-1)^k{\cal
B}_k\left(\Pi^{(n-1)}\right)^k\right]_{[n+1]}
\eeq
\beq\label{deftildePi2}
\tilde \Pi_2^{(n)}=\left[\Lambda\right]_{[n+1]}\frac{{\cal B}_0}{2}+
\left[\delta m^2\right]_{[n+1]}
\eeq
so that $\Pi^{(n)}=\tilde \Pi_0^{(n)}+\tilde \Pi_2^{(n)}$. 

From the previous analysis, we expect the new
subdivergences to occur in the term with $k=1$. However, we have to make sure that subdivergences introduced by
the factor 
$\Lambda$ in front of otherwise finite integrals do cancel.
 In order to do so, we  use Eq.~(\ref{BSitern}) to
express $\Lambda$ in terms of $\Gamma_0$ within Eq.~(\ref{deftildePi0}),
and obtain:
\beq
\tilde \Pi_0^{(n)}&=&\left[\frac{\Gamma_0}{2}\sum_{k=1}^n(-1)^k{\cal
B}_k\left(\Pi^{(n-1)}\right)^k+
\frac{\Gamma_0}{2}{\cal B}_1\tilde \Pi_0^{(n-1)}\right]_{[n+1]}\nonumber\\
&=&\left[\frac{\Gamma_0}{2}\sum_{k=2}^n(-1)^k{\cal B}_k\left(\Pi^{(n-1)}\right)^k+
\frac{\Gamma_0}{2}{\cal B}_1\left(\tilde \Pi_0^{(n-1)}-\Pi^{(n-1)}\right)\right]_{[n+1]}.
\eeq
The first term is now finite. The divergences have been isolated in the last two terms where we can
identify  $\tilde \Pi_0^{(n-1)}-\Pi^{(n-1)}=-\tilde \Pi_2^{(n-1)}$. One then gets for $\Pi^{(n)}=\tilde
\Pi_0^{(n)}+\tilde \Pi_2^{(n)}$:
\beq\label{eq35a}
\Pi^{(n)}=\left[\frac{\Gamma_0}{2}\sum_{k=2}^n(-1)^k{\cal B}_k\left(\Pi^{(n-1)}\right)^k\right]_{[n+1]}
-\left[\frac{\Gamma_0}{2}{\cal B}_1\tilde \Pi_2^{(n-1)}\right]_{[n+1]}+\tilde \Pi_2^{(n)}.
\eeq
Using the definition (\ref{deftildePi2}) of $\tilde \Pi_2$, one can combine the last two terms as follows
\beq\label{eq35}
\left[\Lambda-\frac{\Gamma_0}{2}{\cal B}_1\Lambda\right]_{[n+1]}\frac{{\cal B}_0}{2}+
\left[\left(  1-\frac{\Gamma_0}{2}{\cal B}_1\right)\delta m^2\right]_{[n+1]}.
\eeq
But the divergent terms in this expression are those of the right hand side of Eq.~(\ref{m2dPidm2a1}) (see the discussion after Eq.~(\ref{Piiteration2})). The two
terms in Eq.~(\ref{eq35}) therefore add up to a finite quantity, 
$\left[\Gamma_0\right]_{[n+1]}{b_0}/{2}$. We are  then left with a finite equation  for determining the finite
part of $\Pi^{(n)}$:
\beq\label{eq35b}
\Pi^{(n)}=\left[\frac{\Gamma_0}{2}\sum_{k=2}^n(-1)^k{\cal B}_k\left(\Pi^{(n-1)}\right)^k\right]_{[n+1]}
+\left[\Gamma_0\right]_{[n+1]}\frac{b_0}{2}.
\eeq
This completes the proof that $\Pi^{(n)}$ is made
finite with the counterterms determined in the previous subsection. 

\subsection{Algebraic solution and scheme dependence\label{sec:tadpolealgebraic}}

We now show that is is possible to obtain the previous solutions of both the Bethe-Salpeter equation and
the gap equation by doing simple algebraic manipulations on the equations themselves, without going explicitly
through the iterations.   We shall call $M$ the physical mass, given by $M^2=m^2+\Pi$. Both $M$ and $\Gamma$ are
physical quantities that are calculated as a function
of the renormalized mass and coupling constant, $m$ and $\lambda$, using the 
self-consistent scheme.

First we note that the renormalized coupling constant $\lambda$ is related to
the bare one $\lambda_B=\mu^{2\epsilon}\Lambda$ by
\beq\label{renormg}
\frac{1}{\lambda}=\frac{1}{\Lambda}+\frac{a_1}{2\epsilon}=\frac{1}{\Lambda}+\frac{1}{32\pi^2\epsilon}\,.
\eeq
This follows immediately form Eqs.~(\ref{BStadbis}) and (\ref{BStadter}). It can be also obtained directly from the Bethe Salpeter equation (\ref{BStad0}),  observing that this equation is finite when expressed in terms of the renormalized coupling constant defined in  Eq.~(\ref{renormg}).   Since $\Lambda=\lambda+\delta \lambda$,
Eq.~(\ref{renormg}) is equivalent to
\beq\label{deltalambda33}
\delta \lambda=\frac{\lambda^2
a_1}{2\epsilon}\,\frac{1}{1-\frac{\lambda a_1}{2\epsilon}}=\frac{\lambda^2}{32
\pi^2\epsilon}\,\frac{1}{1-\frac{\lambda}{32\pi^2\epsilon}},
\eeq
which corresponds indeed to the series of counterterms that 
we have obtained via the diagrammatic analysis (cf. Eq.~(\ref{deltalambdak}). 

Using the fact that  $\lambda_B=\mu^{2\epsilon}\Lambda$ is independent
of $\mu$,  one easily obtains that $\lambda(\mu)$ is running with $\mu$ according to (we take the limit $\epsilon\rightarrow 0$)
\beq\label{betatad}
\frac{d\lambda(\mu)}{d\ln\mu}=\frac{\lambda^2}{16\pi^2}\,.
\eeq
Note that this relation is ``exact'' in the sense that it represents faithfully the relation
between $\lambda_B$ and $\lambda$, to all orders in $\lambda$, within the class of diagrams
which are taken into account in the present approximation. In this respect, it is instructive to
verify how this relation holds as one iterates the Bethe-Salpeter equation in
Sect.~\ref{sec:BStadpole}: By demanding  that the solution $\Gamma_0^{(n)}$ be
independent of
$\mu$ at iteration $n$ (at fixed
$m^2$), the corrections  to Eq.~(\ref{betatad})  are pushed to order $\lambda^{n+2}$;
that is, potential corrections of order $\lambda^{n+1}$ cancel out, in very much the same way
as subdivergences do. Now, while exact in the sense specified above, the
$\beta$ function in  Eq.~(\ref{betatad}) represents only one third of the perturbative one-loop 
$\beta$--function. This is so because, as already mentioned, the resummation performed by the
one-loop gap equation involves the one-loop vertex correction in one channel only, namely the
$t$--channel. The remaining two thirds of the one-loop
$\beta$--function will be recovered after taking the 3-loop skeleton diagram into account (see
next section).

Written as a
function of the renormalized coupling constant, the  renormalized 4-point function takes the form
\beq\label{Gamma0}
\Gamma=\frac{\lambda}{ 1- \frac{\lambda}{32\pi^2}\ln\frac{M^2}{\bar\mu^2} }.
\eeq
Thus $\Gamma$ is nothing but the running coupling constant at the scale $\bar\mu =M$. One can verify  that $\Gamma$ is independent of $\mu$ if $M$ is, which we shall verify soon. 
Note that 
  $\lambda(\mu)$ diverges when $\ln(\bar\mu^2/M^2)=32\pi^2/\Gamma$. The appearance of this  ``Landau pole'' at large scales
is a well known feature of $\phi^4$ field theory in  four dimensions. Another manifestation of the
same difficulty is visible on Eq.~(\ref{deltalambda33}):  when $\epsilon\to 0$, $\delta\lambda\to
-\lambda$ or $\lambda_B\to 0$,  reflecting the so-called ``triviality'' of the theory. 

Knowing $\Gamma$, one could   easily get a finite equation for $\del\Pi/\del m^2$. This is obtained from
Eqs.~(\ref{delPtad}) and (\ref{deltam2Lambda}):
\beq
\frac{\del \Pi}{\del m^2}=-\frac{1}{2}\Gamma b_1(M^2).
\eeq
However, it is not easy to exploit this equation to obtain a finite equation for $\Pi$ itself,
by using methods that we could export to the general case. We shall therefore base our
construction of a finite gap equation on the iterative solution obtained in the previous
subsection. 

There are two observations which will guide this construction. The first observation is that, in   the iterative solution,
 the cancellation of ultraviolet divergences only takes place up to some order dictated by the number of
iterations that have been completed: at iteration $n$ new divergences appear, of order $\lambda^{n+1}$, and 
some are ignored because they are of   order higher than 
$\lambda^{n+1}$ (these high order contributions reenter the calculation as countertems needed to cancel subdivergences in the
following iterations). Thus, in an explicit algebraic solution, one should expect the cancellation of divergences to
occur only for the solution of the gap equation, which effectively sums up all the iterations. The second observation
concerns the particular role of the term linear in $\Pi$ in the expansion of the right hand side of the gap equation in
powers of $\Pi$ (see Eq.~(\ref{expansiongapeq})). As we have seen, the new divergences in the gap equation at each new
iteration come  from diagrams which are obtained by iterating just this term  (e.g. the towers of bubbles in
Fig.~\ref{fig:3loop}). Terms with several insertions of $\Pi$ on the same propagator are finite (assuming that such
insertions have been made finite in previous iterations). With these observations in mind, we now proceed to a construction
of the algebraic solution which will follow closely what has been done in the previous subsection.

We start by expanding the
self-consistent propagator $D$ around the free propagator $D_0$ 
and set (see Eq.~(\ref{expansiongapeq})):
\beq\label{expansionD1}
D(P)=D_0+\delta D= D_0(P)+D_0(P)(-\Pi)D_0(P)+D_r(P)
\eeq
where we have singled out the term linear in $\Pi$, and   $D_r(P)$ behaves as  $\Pi^2/P^6$ at large $P$, so that the integral $\int_P 
D_r(P)$ is finite. Then we generalize the definitions (\ref{deftildePi0}) and
(\ref{deftildePi2}) and set:
\beq\label{ndeftildePi0}
\tilde \Pi_0=\frac{\Lambda}{2} \int_P \delta D(P),
\eeq
and 
\beq\label{ndeftildePi2}
\tilde \Pi_2=\frac{\Lambda}{2} \int_P D_0(P)+\delta m^2.
\eeq
In these notations, the gap equation reads
\beq\label{gap1}
\Pi=\tilde \Pi_0+\tilde \Pi_2,
\eeq
where only $\tilde \Pi_0$ depends on the solution $\Pi$ ($\tilde \Pi_2$ is entirely determined by
$D_0$, as is obvious from Eq.~(\ref{ndeftildePi2}) and the fact that  the
counterterms $\delta \lambda$ (in $\Lambda=\lambda+\delta\lambda$) and $\delta m^2$ are calculated entirely from $D_0$).

 Next, one uses the Bethe-Salpeter
equation to eliminate $\Lambda$ in favor of $\Gamma_0$ in the defining equation for $\tilde\Pi_0$ (Eq.~(\ref{ndeftildePi0}))
and get
\beq\label{tildePi0tad}
\tilde \Pi_0&=&\frac{\Lambda}{2}\int_P \delta D(P)\nonumber\\
&=&\frac{\Gamma_0}{2}\int_P \delta
D(P)+\frac{\Gamma_0}{2}\, \tilde\Pi_0\int_Q D_0^2(Q)\nonumber\\
&=&\frac{\Gamma_0}{2}\int_P 
D_r(P)+\frac{\Gamma_0}{2}\, (\tilde\Pi_0-\Pi)\int_Q D_0^2(Q).
\eeq
At this point we observe that when $\Pi$ is solution of the gap equation (\ref{gap1}), $\tilde\Pi_0-\Pi=-\tilde\Pi_2$, so
that the divergence of $\tilde \Pi_0$ (isolated in the second term of  Eq.~(\ref{tildePi0tad})) is independent of 
$\Pi$: this is the main achievement of the elimination of vertex subdivergences. 

The  gap equation then reads:
\beq
\Pi=\frac{\Gamma_0}{2}\int_P D_r(P)+\tilde\Pi_2\left(1-\frac{\Gamma_0}{2}\int_Q D_0^2(Q)\right).
\eeq
The first term, which involves $\Pi$ (through $D_r$) is finite. 
In the second term, we note that the
factor multiplying
$\tilde\Pi_2$ is
$\Gamma_0/\Lambda$: this follows from the Bethe-Salpeter with propagators $D_0$. Now
\beq
\frac{\Gamma_0}{\Lambda}\tilde \Pi_2=\frac{\Gamma_0}{\Lambda}\left(\frac{\Lambda}{2}{\cal
B}_0+\delta m^2\right)= \frac{\Gamma_0}{2} b_0(m^2),
\eeq
where $b_0$ is the finite part of ${\cal B}_0$, and we have set 
(see Eq.~(\ref{deltam2Lambda}))
\beq\label{deltam2a}
\delta
m^2=\frac{\Lambda m^2}{32\pi^2\epsilon}.
\eeq 
The gap equation takes then the form: 
\beq\label{Pifinite}
\Pi &=&
\frac{\Gamma_0}{2}\int_P D_r(P)\,+\,\frac{\Gamma_0}{32\pi^2}\,m^2\,
\left(\ln \frac{m^2}{\bar\mu^2}-1\right)\nonumber\\
&=&\frac{\Gamma_0}{32\pi^2}\left[(m^2+\Pi)\ln \frac{m^2+\Pi}{m^2}-\Pi\right]+\,\frac{\Gamma_0}{32\pi^2}\,m^2\,
\left(\ln \frac{m^2}{\bar\mu^2}-1\right)
\eeq
which is now a finite equation. It can also be written in terms of the renormalized coupling constant:
\beq\label{gaptad2}
\frac{\Pi}{\lambda}\,= \frac{m^2+\Pi}{32\pi^2}\left(\ln\frac{m^2+\Pi}{\bar\mu^2}
-1\right)=\frac{1}{2} b_0(M^2),\eeq
or in terms of the physical  4-point function
$\Gamma$:
\beq\label{Mtot}
M^2= \Gamma\left[\frac{m^2}{\lambda} - \frac{M^2}{32\pi^2}\right] .
\eeq

This last form, in which all explicit $\mu$-dependence has disappeared, makes it obvious that $M$ is independent of $\mu$,
because
$m^2/\lambda$ is independent of
$\mu$.  Indeed,  $m_B^2/\lambda_B
= (m^2+\delta m^2)/(\mu^{2\epsilon}\Lambda)$ has a finite limit when
$\epsilon\to 0$, equal to $m^2/\lambda$. Since $m_B^2/\lambda_B$   is manifestly
independent of $\mu$, so is 
therefore $m^2/\lambda$, which further
implies that $m$ runs with $\mu$ in the same way as $\lambda$ 
does (i.e., according to Eq.~(\ref{betatad})).

As made obvious by
the previous discussion, the gap equation can be written either in terms of $\Gamma_0$ or in terms of
$\Gamma$. This reflects a property that we have repeatedly
emphasized, namely that the divergent parts of the counterterms do not depend on whether $D$ or $D_0$ is used in their
calculations. We can exploit this freedom in rearranging the expansion of Eq.~(\ref{expansionD1}) in the following way:
\beq\label{expansionD2}
D(P)=D_0+\delta D= D_0(P)+D(P)(-\Pi)D(P)+D_r'(P)
\eeq
where $D_r'(P)$ behaves like $1/P^6$, so that the integral $\int_P 
D_r'(P)$ is finite. Then the full propagator 
$D$ will naturally appear in the Bethe-Salpeter equation for the vertex
counterterms of the gap equation. 
We can verify that all the steps leading from Eq.~(\ref{expansionD1}) to the final gap
equation  (\ref{Mtot}) still hold, with the appropriate replacements of
$\Gamma_0$ by
$\Gamma$, $D_0$ by $D$, and 
$D_r$ by
$D'_r$ (note that
$\tilde
\Pi_2$ is not affected). The gap equation (\ref{Mtot}) remains of course unaltered. 
The advantage of this new expansion 
becomes apparent when studying the massless limit to which we proceed 
in the next subsection. 

Finally, we note that we could have used a
different renormalisation scheme such as the mass--shell
renormalization, in which we assume that $m^2$ is the physical mass,
i.e., $m^2=M^2$. 
In this scheme  $\Pi=0$, and the mass counterterm is given by
\beq\label{dmI}
\delta m^2=-\frac{\lambda_B}{2}\int_P\frac{1}{P^2+m^2}\,.
\eeq
 The coupling constant counterterm can  still be chosen as
in eq.~(\ref{renormg}), and we have 
$\Gamma=\Gamma_0=\lambda(\bar\mu=m)$. This mass--shell scheme,
 which makes the gap equation 
trivial at zero temperature, simplifies it at finite temperature, 
as we shall see in the next subsection.

\subsection{Finite temperature and massless limit\label{sec:masslesstadpole}}

The finite--temperature version of
the gap equation (\ref{Pitad}) reads:
\beq\label{PitadT}
\Pi = \frac{\Lambda}{2}\mu^{2\epsilon}\int_P D(P)\,+\,
\frac{\Lambda}{2}\int_p\,\frac{n(\varepsilon_p)}{\varepsilon_p}\, +\delta m^2,
\eeq
where $n(\varepsilon_p)=1/({\rm e}^{\beta \varepsilon_p}-1)$,  
$\varepsilon_p=\sqrt{p^2+M^2}$ and $\int_p$ denotes a 3-dimensional momentum integral. The first
integral
$\int_P D(P)$ is to be understood as before as a $d$-dimensional Euclidean integral, but 
$D$ now depends on the temperature. 

One could  naively fear that temperature dependent divergences could appear in the calculation of $\Pi$, as for instance in
the second term of Eq.~(\ref{PitadT}), due to the presence of the factor $\Lambda$. This however
does not happen if one carefully eliminates
 all the subdivergences which emerge when solving the gap equation iteratively. 
To understand this, it is enough to consider the first two iterations. 

The zeroth iteration is given by Eq.~(\ref{Piiteration0}), with $b_0$ replaced by
$b_0+b_0^T$. Clearly the countertem (\ref{deltam0}) makes $\Pi^{(0)}$ finite. The
first iteration is given by Eq.~(\ref{deltam2}), with the finite parts of the
integrals including their finite temperature contributions. Potential difficulties
could arise for instance from diagram (a) in Fig.~\ref{fig:2loop} in which the temperature dependent piece
of the upper loop multiplies the divergent lower loop. But it is easy to verify
that this does not happen: indeed the divergence of the lower loop is cancelled by
the coupling constant counterterm of diagram (b). Similarly, the contribution proportional to the mass counterterm $\delta
m_0^2$ multiplying the finite temperature part of the loop integral cancels by the same mechanism as at zero temperature.
Thus, for the same reason that once subdivergences are eliminated, there remains no finite part coming from some
subintegrations multiplying divergent part coming from other subintegrations, there does not remain either finite
temperature part multiplying divergent integrals or infinite counterterms.  This analysis can be generalized, by going
through the same steps as in  subsection \ref{sec:itsolBStadpole},  in order to verify that, iteration by iteration, all
potentially temperature dependent infinities actually cancel.

To obtain this result in a more explicit fashion, we observe that the term with a statistical
factor in Eq.~(\ref{PitadT}) can be viewed as a new contribution to
$D_r$, which is then easy to incorporate  in the renormalized gap equation  
 (\ref{gaptad2}):
\beq\label{gaptad2a}
\frac{\Pi}{\lambda}\,= \frac{m^2+\Pi}{32\pi^2}\left(\ln\frac{m^2+\Pi}{\bar\mu^2}
-1\right)+\frac{1}{2}\int_p\,\frac{n(\varepsilon_p)}{\varepsilon_p}.\eeq
In the mass-shell renormalization scheme, we get instead
\beq\label{gaptad2ab}
\Pi=\frac{\Gamma_0}{2}\int_P D_r(P)+\frac{\Gamma_0}{2}\int_p\,\frac{n(\varepsilon_p)}{\varepsilon_p},
\eeq
which implies that $\Pi\to 0$ as $T\to 0$.

Consider now the massless limit $m\to 0$. In this limit, many of
the intermediate manipulations based on the expansion (\ref{expansionD1})
of the full propagator $D$ around the free propagator $D_0$ become 
meaningless because of infrared problems. For instance, the 
Bethe-Salpeter equation with propagator $D_0$, Eq.~(\ref{BStad0}), is
ill-defined when $m\to 0$, and therefore so is also the 4-point
function $\Gamma_0$. More generally,  all the integrals ${\cal B}_k$ 
with $k\ge 1$ become infrared divergent when $m=0$. In particular,
one cannot take the limit $m\to 0$ on the gap equation written
in the form (\ref{gaptad2ab}) (or (\ref{Pifinite}),
since both $\Gamma_0$ and the integral 
involving $D_r$ develop infrared divergences in this limit. 
Still, the gap equation {\it has} a well defined massless limit, 
as can be seen by taking this limit at the level of Eq.~(\ref{gaptad2a}). 
This gives:
\beq\label{gaptad3}
{\Pi}\,=\,\frac{\lambda}{32\pi^2}\,\Pi\left(\ln\frac{\Pi}{\bar\mu^2}
-1\right)\,+\,
\frac{\lambda}{2}\int_p\,\frac{n(\varepsilon_p)}{\varepsilon_p},\eeq
which shows that,  at finite temperature, the finite self-energy $\Pi$ 
provides an infrared cut-off, allowing the limit $m\to 0$ to be taken.
This limit can also be written in terms of the  
full 4-point function $\Gamma$:
\beq\label{tad-massless}
{\Pi}\,=\, - \frac{\Gamma}{32\pi^2}\,\Pi +
\frac{\Gamma}{2}\int_p\,\frac{n(\varepsilon_p)}{\varepsilon_p}.\eeq
Here, $\Gamma$ is given by Eq.~(\ref{Gamma0}) with $M^2=\Pi$.
Thus  $\Gamma$  itself depends upon the temperature.
From either Eq.~(\ref{gaptad3}) or Eq.~(\ref{tad-massless}) one can
check that, at zero temperature, $\Pi=0$ in this massless case.
As we shall see in Sect. IV, the general approach to renormalization
in the massless case is based on the modified expansion in
Eq.~(\ref{expansionD2}), which takes advantage of the presence
of the finite-temperature self-energy $\Pi$ as a natural infrared cut-off.

\section{The three-loop skeleton}
\setcounter{equation}{0}

Many of the developments of the previous section can be carried over to the discussion of the three-loop skeleton. In
particular, the determination of the coupling constant and mass counterterms will go through the solution of the
Bethe-Salpeter equation and the equation for $\del\Pi/\del m^2$, with little structural changes with respect to the two-loop
examples. The essential new aspect of the
three-loop example is the presence of the field renormalization constant $\delta Z$, and the fact that the self-energy
contributes to  the asymptotic behaviour of the propagator. 

 The three loop skeleton for $\Phi$ is 
displayed in Fig.~\ref{skeletons}, and its contribution reads:
\beq
\Phi=-\frac{\lambda^2}{48}\int_P\int_Q\int_R D(P) D(Q) D(R) D(K+P+Q+R)+\frac{\delta\lambda}{8}\left(\int_P D(P) 
\right)^2.
\eeq
We have added here in $\Phi$ a counterterm which has the structure of 
the two-loop skeleton; this term is  needed to cancel  subdivergences, as we shall see.
 Such a
term would be automatically present if we had included in $\Phi$ both the three-loop
and the two-loop skeletons.
The self-energy  follows then from Eq.~(\ref{eq:gap}):
\begin{equation}\label{eq:gapA}
\Pi(K)\,=\,-\frac{\lambda^2}{6}\int_P\int_Q D(P)D(Q)D(K+P+Q)+
\,\frac{\delta \lambda}{2}\int_P D(P)+\delta m^2+K^2\delta Z, 
\end{equation}
and the  kernel of the Bethe-Salpeter equation from
Eqs.~(\ref{Lambda0}):
\beq\label{lambdasunset0}
\Lambda(P,K)=-\,\lambda^2\int_Q D(Q)D(K+P+Q)+\delta \lambda.
\eeq

It is important to observe that the vertices of the three-loop contribution in $\Phi$ do not carry coupling
constant counterterms.   We shall see indeed  that, as as we proceed to solving 
the gap equation through iterations, we do not generate structures which
would renormalize these vertices. 
Such structures would appear only at the 4-loop order in the skeleton expansion. This is a particular
illustration of a general feature of $\Phi$-derivable approximations that will be discussed in the next section.

As we
indicated in Sect.~\ref{sec:formalism}, and verified explicitly in the two loop example of
Sect.~\ref{sec:twoloop}, the subdivergences associated with vertex and mass renormalizations can be
determined from the analysis of the Bethe-Salpeter equation and of the equation for the derivative of the
self-energy with respect to $m^2$. We shall then go through such an analysis. But first
 we discuss in some details the first orders in perturbation theory obtained from 
the first two iterations of the gap equation.

\subsection{Perturbation theory\label{sec:perturbation}}

\begin{figure}[htb]
\epsfysize=3.cm
\centerline{\epsffile{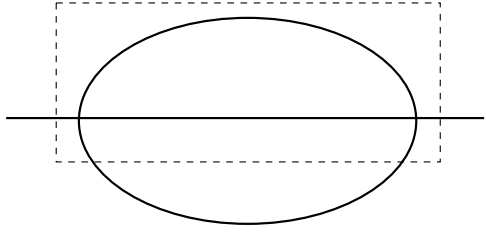}}
\vspace{2mm}
\caption[a]{ The dotted line box encloses the subgraph whose divergence is that of the 4-point function
$\Lambda_0$.
\label{SSbox}}
\end{figure}

The zeroth order iteration  $\Pi^{(0)}(K)$ is obtained by replacing  $D$  by
$D_0=(P^2+m^2)^{-1}$ in the r.h.s. of eq.~(\ref{eq:gapA}), and coincides with the standard
2--loop contribution of perturbation theory. 
In dimensional regularization, the divergent part of the diagram is of the form \cite{Ramond:yd}:
\beq\label{SSdiv}
\Pi_{\mbox{\tiny div}}^{(0)}(K)
=\frac{\lambda^2}{(16\pi^2)^2}\left\{ \frac{K^2}{24\epsilon}+m^2\left[\frac{1}{4\epsilon^2}
+\frac{3}{4\epsilon}+\frac{1}{2\epsilon}\ln\frac{\bar \mu^2}{m^2}\right]\right\}.
\eeq
Hidden in these  various terms there are three distinct divergences. One is a subdivergence which is
removed by a coupling constant counterterm. It will be dealt with shortly. The other
two divergences are global divergences corresponding respectively to a mass
divergence and a field normalisation divergence. These will be removed by adjusting
respectively $\delta m^2$ and $\delta Z$. The contribution of
$\delta Z$ is easy to identify since it is associated with the unique divergent term
proportional to $K^2$. Using minimal subtraction we get the following
lowest-order contribution $\delta Z_0$ to $\delta Z$ :
\beq
\delta Z_0=-\frac{\lambda^2}{(16\pi^2)^2}\frac{1}{24\epsilon}.
\eeq
The adjustment of $\delta m^2$ can be made
only after the proper elimination of the vertex subdivergence. 
 The latter singles out one
line in the sunset diagram, namely, that line along which the momentum
is kept finite, while the other loop momentum is sent to infinity. In fact, the
corresponding subgraph  can be identified with
$\Lambda_0$ (see Fig.~\ref{SSbox}), the leading order contribution to $\Lambda$, 
to within the counterterm $\delta\lambda$; that is: 
\beq\label{Lambda0calcul}
\Lambda_0(P,K)-\delta\lambda=-\frac{\lambda^2}{16\pi^2\epsilon}+
\frac{\lambda^2}{16\pi^2}\int_0^1{\rm d}x
\ln\frac{|m^2+x(1-x)(P+K)^2|}{\bar\mu^2}.
\eeq 

\vspace{2mm}

Let us call $\delta\lambda_0$ the value of $\delta\lambda$  that renders $\Lambda_0$ finite.  Minimal
subtraction gives:
\beq\label{DeltaLambda0}
\delta \lambda_0\,=\,\frac{\lambda^2}{16\pi^2\epsilon}\,.
\eeq
{Note that this contribution, when added to that of  Eq.~(\ref{renormg}),
reconstructs the full order $\lambda^2$
counterterm ($3\lambda^2/32\pi^2\epsilon$).
The counterterm (\ref{DeltaLambda0}) generates an order $\lambda^2$ contribution to $\Pi$ of the form:
\beq\label{subdivS0}
\frac{\delta\lambda_0}{2} {\cal B}_0(m^2)=  \frac{\delta\lambda_0}{2}\left( \frac{a_0}{
\epsilon}+b_0\right) ,
\eeq
which eliminates the subdivergence in $\Pi$, Eq.~(\ref{SSdiv}), and the
logarithmic term there (via  
$(\delta \lambda_0/2) b_0$ involving the finite part of ${\cal B}_0$ 
in Eq.~(\ref{subdivS0})). There remains
then only a  global, mass  divergence, which is removed by
appropriately tuning the counterterm 
$\delta m^2$. Minimal subtraction gives:
\beq\label{deltam2sunsetpt}
\delta m^2_0=\left(\frac{\lambda
m}{32\pi^2}\right)^2\left(\frac{1}{\epsilon^2}-\frac{1}{\epsilon}\right).
\eeq

Note that neither $\delta \lambda_0$, $\delta Z_0$ nor $\delta m^2_0/m^2$ depend on $m$: 
these counterterms depend only on the
asymptotic form of the propagator. Note also that 
$\delta Z_0$ could be obtained by analyzing the
divergence which remains after differentiating $\Pi^{(0)}$ with respect to $K^2$. 
Since the coupling constant counterterm
is independent of momentum, it does not contribute to the latter derivative, and the same applies to
the mass counterterm.  This explains why the determination of $\delta Z_0$ can be made
independently of the elimination of the vertex and mass subdivergences. This property persists in higher orders, as we shall see in Sect.~\ref{sec:deltaZ}

\vspace{2mm}
\begin{figure}[htb]
\epsfysize=9.5cm
\centerline{\epsffile{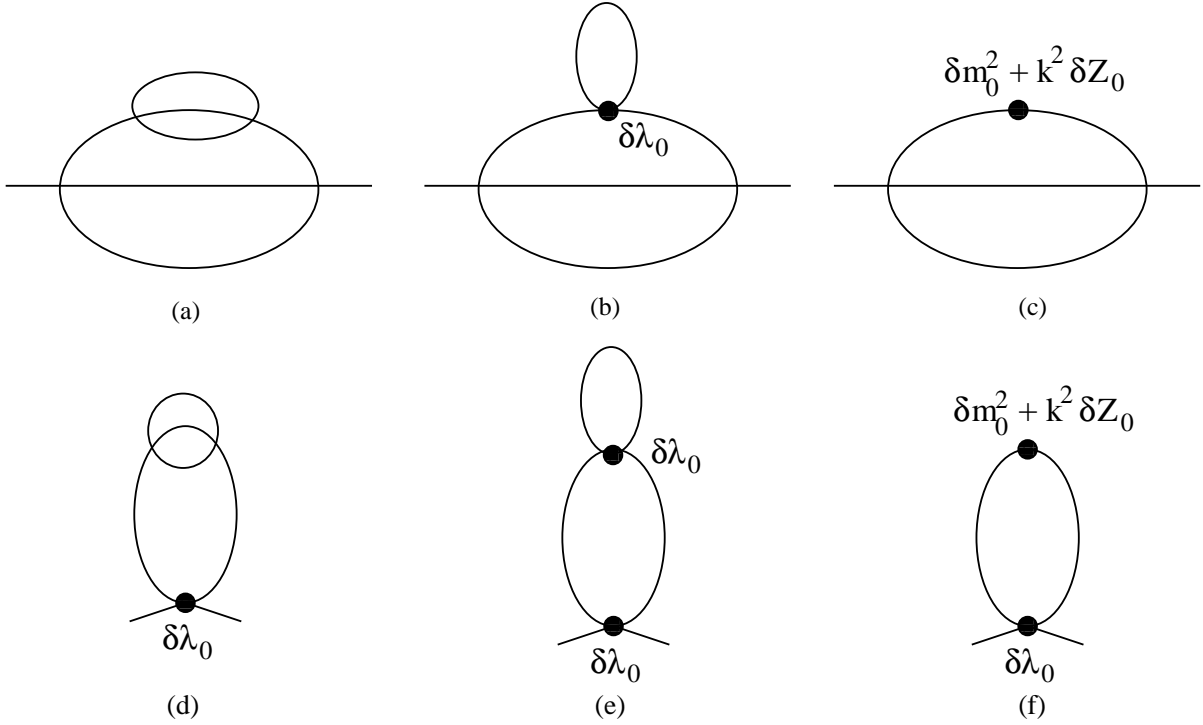}}
\vspace{2mm}
\caption[a]{The diagrams contributing to $\Pi^{(1)}$, the solution
to the gap equation to order $\lambda^4$. Diagram (a) is the original diagram with one insertion. All other
diagrams involve counterterms which remove the various subdivergences. \label{SS1it}}
\end{figure}

For later purposes, we note that 
the asymptotic behaviour of $\Lambda_0$  (after renormalization) is of the form 
$\Lambda_0(P,K)\sim\ln(P+K)^2/\mu^2$ at large momenta, 
in agreement with Weinberg's theorem \cite{Weinberg,Collins:xc}
 (that $\Lambda_0(P,K)$ 
is a function of $P+K$ can be seen directly
on Eqs.~(\ref{Lambda0calcul}) or (\ref{lambdasunset0})). As for $\Pi^{(0)}(K)$, its dominant contribution at
large $K$ goes as $K^2 \ln K^2$ (this may be obtained by calculating the finite part of $\Pi^{(0)}(K)$ for
$m=0$).

Consider now the first iteration in which we keep only the contributions of order up to 
$\lambda^4$, and call $\Pi^{(1)}$ the
corresponding value of $\Pi$. The associated diagrams are shown in Fig.~\ref{SS1it}. Not
represented there are  the diagrams involving the new counterterms that need to be adjusted
at this iteration,  namely,
$\delta \lambda_1$, $\delta m^2_1$,
and $\delta Z_1$.

Let us focus on the new vertex subdivergences. The corresponding diagrams
are displayed in Fig.~\ref{2STa}. The subgraph within the box in  Fig. \ref{2STa}.a is logarithmically
divergent because the inserted self-energy grows like
$K^2\ln K^2$ at large $K$, as mentioned earlier;
thus, this insertion does not change the degree of divergence of the diagram,
the factor $K^2\ln K^2$ cancelling  the $1/K^2$ coming from the extra propagator.  
 The subdivergence in  Fig. \ref{2STa}.a (together with the corresponding counterterms from Figs.~\ref{SS1it}.b 
and \ref{SS1it}.c)   can be recognized as the contribution of order $\lambda^4$
to the (global) divergence of the integral in Eq.~(\ref{lambdasunset0}) for
the kernel $\Lambda$ of the Bethe-Salpeter equation. In other terms, the subgraph
singled out in Fig. \ref{2STa}.a represents the first correction to the zeroth order
kernel $\Lambda_0$ in Eq.~(\ref{Lambda0calcul}), and its divergence can be absorbed in a correction
$\delta\lambda_0^{(1)}$ of order $\lambda^4$ to $\delta\lambda_0$. This type of correction is a new feature
which did not appear in the two-loop example: it is related to the fact that the insertion of a self-energy affects the asymptotic behaviour of the propagator. However, the other 4-point divergence, corresponding to 
\vspace{2mm}
\begin{figure}[htb]
\epsfysize=4.cm
\centerline{\epsffile{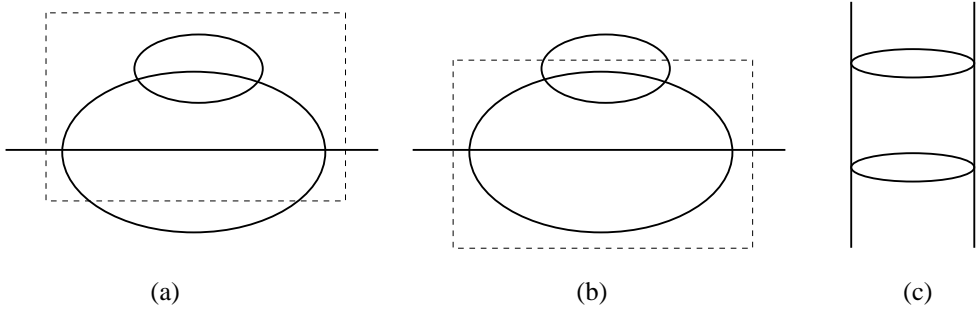}}
\vspace{2mm}
\caption[a]{Vertex subdivergences in the diagrams contributing to $\Pi^{(1)}$.
The subdivergence represented by the dashed-line box in figure (b) is that of the
4--point function depicted in figure (c), and is removed by a counterterm 
determined by the Bethe-Salpeter equation. \label{2STa}}
\end{figure}
the subgraph in the box in Fig.~\ref{2STa}.b is analogous to  those encountered in the two-loop example. It is
the  divergence of the 4--point function shown in Fig.~\ref{2STa}.c, obtained as the first
iteration of the  Bethe-Salpeter equation with the leading order kernel $\Lambda_0$. The corresponding
divergence  is absorbed in the counterterm $\delta\lambda_1$ (after subdivergences are eliminated by
counterterms such as those in Figs.~\ref{SS1it}.b, d and e).

To summarize,  at the first iteration, and order $\lambda^4$, the vertex counterterm 
receives two contributions: one, which we call $\delta\lambda_0^{(1)}$, removes the
order--$\lambda^4$ divergence from the kernel $\Lambda$, the other,
which we call $\delta\lambda_1$,  renormalizes the first iteration of the Bethe-Salpeter equation. Once the vertex subdivergences are
removed, the mass counterterm $\delta m_1^2$ can be chosen so as to remove the remaining 
mass divergences from $\Pi^{(1)}$. Finally, $\delta Z_1$ can be obtained by analyzing the
divergence which remains after differentiating $\Pi^{(1)}$ with respect to $K^2$.
The same argument that we used at leading order shows that this can be computed independently of the
vertex and mass renormalizations, that is, $\delta Z_1$ can be obtained by extracting the
divergence proportional to $K^2$ in the sum of the two diagrams (a) and (c) of
Fig.~\ref{SS1it} (with only $K^2\delta Z_0$ in diagam (c)).

In order to obtain a more systematic determination of the coupling constant and mass counterterms, we turn now
to the analysis of the Bethe-Salpeter equation and the equation for $\del\Pi/\del m^2$. We shall return to $\delta Z$ in Sec.~\ref{sec:deltaZ}.

\subsection{The Bethe-Salpeter equation}

As we did in the previous section, 
we shall consider the solution obtained by iterations, using first in 
the integral of Eq.~(\ref{BSdef}) the free
propagator $D_0$. We shall, as before, call  $\Gamma_0$ the corresponding solution, $\Lambda_0$ the
associated kernel, and construct
$\Gamma_0$ as a formal series  by solving the Bethe-Salpeter equation  by iterations. 
We have: 
\beq\label{G0a}
\Gamma_0(P,K)&=&\Lambda_0(P,K)-\frac{1}{2}\int_Q \Lambda_0(P,Q)D_0^2(Q)\Gamma_0(Q,K)
\nonumber\\
&=&\Lambda_0(P,K)-\frac{1}{2}\int_Q \Gamma_0(P,Q)D_0^2(Q)\Lambda_0(Q,K).
\eeq

In leading order, and before renormalization, $\Gamma^{(0)}_0$ is  equal to $\Lambda_0$, which is  divergent. This  divergence has been dealt with in
Sect.~\ref{sec:perturbation} where we have determined the counterterm $\delta \lambda_0$ needed to absorb it
($\delta \lambda_0$ is given by Eq.~(\ref{DeltaLambda0})). We shall call $\tilde\Lambda_0$ the finite quantity
\beq\label{lambdasunset}
\tilde\Lambda_0(P,K)\equiv\delta \lambda_0\!-\!\lambda^2\int_Q D_{0}(Q)D_{0}(K+P+Q).
\eeq
Thus, after renormalization,  $\Gamma^{(0)}_0=\tilde\Lambda_0$.
The quantities $\tilde\Lambda_0$ and $\Lambda_0$ differ by the counterterm difference
$\delta\lambda-\delta\lambda_0$ which will be needed to remove further divergences
 appearing as we proceed with the iterative solution of Eq.~(\ref{G0a}).

\begin{figure}[htb]
\begin{center}
\epsfysize=7.cm
\centerline{\epsffile{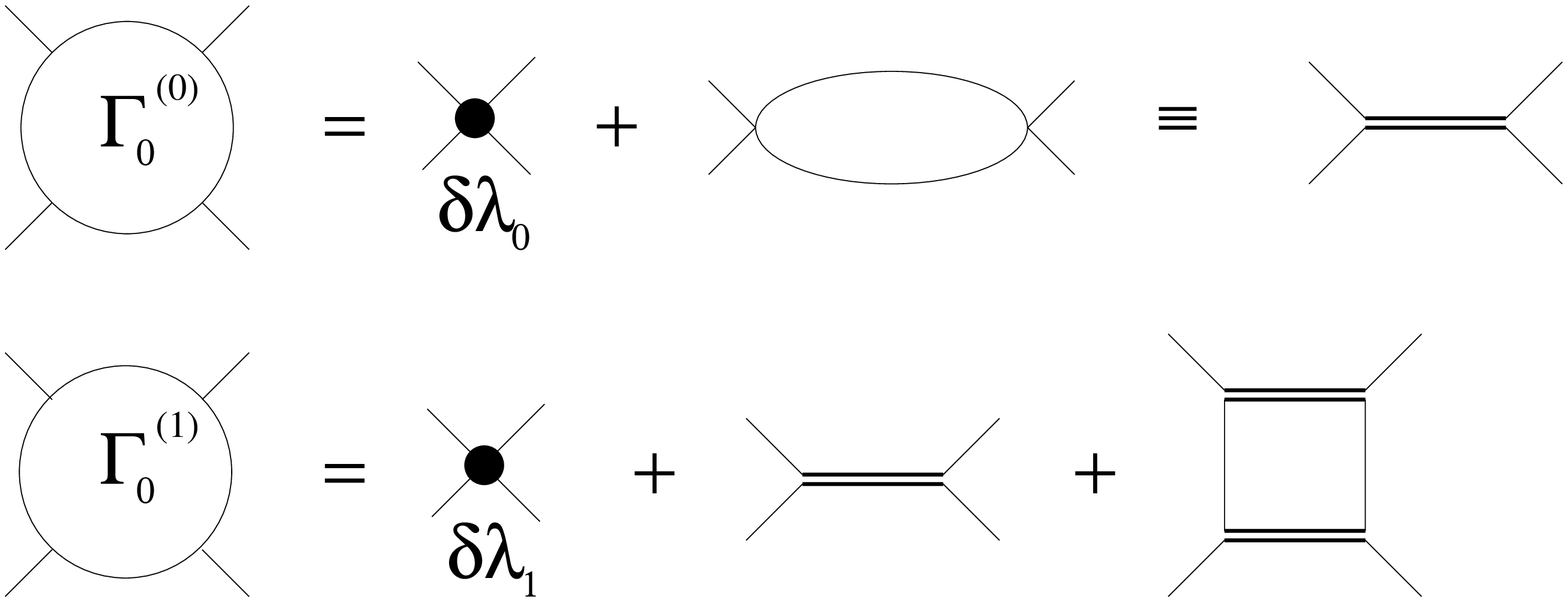}}
\caption{The contributions to $\Gamma_0$
 obtained from the zeroth and first iteration of 
the Bethe-Salpeter equation. The horizontal double bar represents $\tilde\Lambda_0$. The counterterm
$\delta\lambda_1$ absorbs  the divergence of the loop integral in the last diagram 
for $\Gamma_0^{(1)}$ (see Eq.~(\ref{G0a1})).
\label{fig:BSsunset1}}
\end{center}
\end{figure}

 The first iteration reads
\beq\label{G0a1}
\Gamma^{(1)}_0(P,K)=\tilde\Lambda_0(P,K)+
\delta\lambda_1-\frac{1}{2}\int_Q \tilde\Lambda_0(P,Q)D_0^2(Q)\tilde\Lambda_0(Q,K).
\eeq
The right hand side contains no subdivergences: these are eliminated by the counterterm which
makes $\tilde\Lambda_0$ finite. What remains in the integral is then a global divergence which
can be absorbed in $\delta\lambda_1$. That the divergence of the integral in  Eq.~(\ref{G0a1})
 is a local one
can be easily verified by observing that the derivative of this integral with respect to $P_\mu$ leaves a
finite integral. This can be checked by using the explicit form of
 $\tilde\Lambda_0(P,K)$ given in Eq.~(\ref{lambdasunset}).  This implies that the counterterm cannot depend on
$P$. For the same reason, it cannot depend on
$K$. Hence it is a constant, which  can be chosen by minimal subtraction, as we did earlier for
$\delta\lambda_0$, or by imposing a renomalization condition on $\Gamma_0(P,K)$ at some values of the
momenta. We shall implicitly continue using minimal subtraction. A diagrammatic interpretation of the
first two iterations is given in  Fig.~\ref{fig:BSsunset1}.

It is instructive to perform another iteration and get:
\beq\label{G0a3}
\Gamma^{(2)}_0(P,K)&=&\tilde\Lambda_0(P,K)+\delta\lambda_1+\delta\lambda_2\nonumber\\
&-&\frac{1}{2}\int_Q \left[\tilde\Lambda_0(P,Q)+\delta\lambda_1\right]D_0^2(Q)
\left[\tilde\Lambda_0(Q,K)+\delta\lambda_1\right]\nonumber\\
&+&\frac{1}{4}\int_Q\int_R\tilde\Lambda_0(P,Q)D_0^2(Q)
\tilde\Lambda_0(Q,R)D_0^2(R)\tilde\Lambda_0(R,K),
\eeq
where we have dropped terms of order $\lambda^8$ or higher. The contributions of
order $\lambda^6$ are depicted in Fig.~\ref{fig:BSsunset2}. One  can then again proceed to
an analysis of the subdivergences. Note that there are no subdivergence
involving only one line of a $\Lambda_0$ (since opening one line of a $\Lambda_0$ would lead to a convergent
6-point function); in other words the divergent subdiagrams contain complete $\Lambda_0$ 
as subdiagrams \cite{vanHees:2001ik}. Again, one finds that all the
subdivergences are eliminated by the counterterms determined at the previous iteration, namely
$\delta\lambda_0$ and
$\delta\lambda_1$. The remaining global divergence is absorbed in the new local counterterm
$\delta\lambda_2$. That this divergence is a local one can be verified by the same argument as given after
Eq.~(\ref{G0a1}), namely by differentiating with respect to the external momenta and verifying that all the
resulting contributions are finite.

\begin{figure}[htb]
\begin{center}
\epsfysize=3cm
\centerline{\epsffile{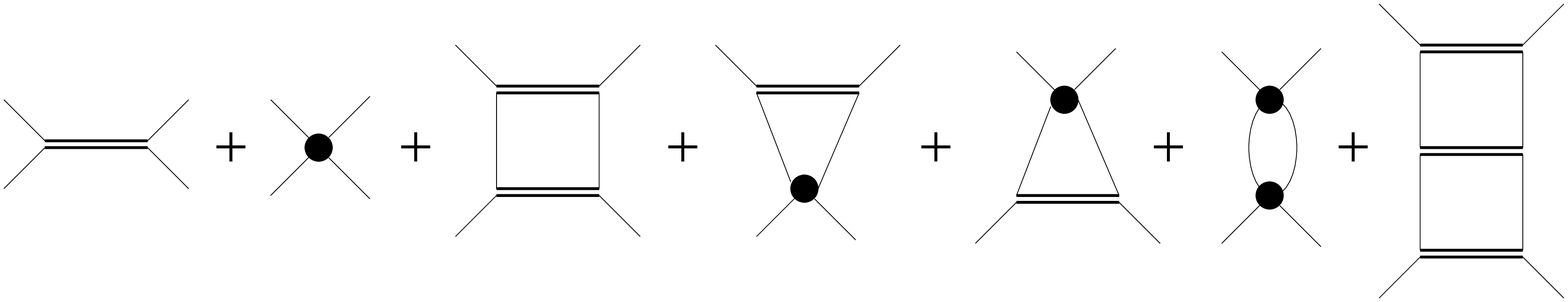}}
\caption{The contributions of order $\lambda^6$ to $\Gamma_0^{(2)}$
 obtained from the second iteration of the Bethe-Salpeter equation, illustrating the
elimination of the subdivergences leading to the determination of the global
counterterm $\delta\lambda_2$.  The kernel
$\tilde\Lambda_0$ (which includes its counterterm
$\delta\lambda_0$) is represented by a horizontal bar, as in Fig.~\ref{fig:BSsunset1}. The black dots represents the counterterm $\delta\lambda_1$, except for the first one on the left which is the sum of $\delta\lambda_1$ and $\delta\lambda_2$. \label{fig:BSsunset2}}
\end{center}
\end{figure}

It is clear that the successive iterations of the Bethe-Salpeter equation will
reproduce a structure analogous to that obtained in Sect.~\ref{sec:itsolBStadpole}
for the two-loop skeleton, with the 
contribution $\Gamma^{(n)}_0$ at iteration $n$ being of order $\lambda^{2n+2}$. 
In particular, the contributions involving
the product of a counterterm $\delta\lambda_k$ by a finite integral 
cancel out (such terms 
appear for instance in the second line of Eq.~(\ref{G0a3})
and compensate contributions of the first line where one of the 
integration variables is kept fixed while the other one is taken to infinity).
This is the kind of cancellations expected in the BPH analysis \cite{Collins:xc},
and guarantees, in
particular, that when the calculation is done at finite temperature, 
there does not remain, at any iteration, terms involving
the product of an infinite counterterm by a finite temperature contribution. 
  
The only significant difference with the 2-loop example discussed in the previous section,  is that the divergences do not
simply factorize because the kernel is momentum dependent, and no simple  expression can be given for
the 
 counterterms. 
Because of this, one may find  useful  to have finite
equations for $\Gamma$ in which those counterterms do not explicitly enter. 
Consider for instance $\Gamma_0(0,P)$. This  can
be obtained from the following finite equation:
\begin{eqnarray}\label{eqG}
\!\!\!\!\Gamma_0(0,P)-\Gamma_0(0,0) &\!=\!&
\Lambda_0(0,P)\!-\!\Lambda_0(0,0)\nonumber\\
&{}& -\,\frac{1}{2}\int_Q
\Gamma_0(0,Q)D_{0}^2(Q)\big\{\Lambda_0(Q,P)\!-\!\Lambda_0(Q,0)\big\}.\,\,\,\,\,
\end{eqnarray}
It is easily verified that  $\Lambda_0(Q,P)-\Lambda_0(Q,0)\sim P/Q$ for $Q^2\gg
P^2$. Since $\Gamma_0(0,Q)$ grows at most logarithmically
 (this follows from Weinberg's theorem \cite{Weinberg,Collins:xc}),  the integral
over
$Q$ in Eq.~(\ref{eqG}) is indeed finite. Thus,
Eq.~(\ref{eqG}) is finite, and can be solved to determine
$\Gamma_0(P,0)$ in terms of $\Gamma_0(0,0)$,  which 
is supposed to be finite and may be chosen to satisfy a renormalization condition. 
Once $\Gamma_0(0,P)$ is known, it may be used to determine the full momentum
dependence of $\Gamma_0(K,P)$. The difference
$\Gamma_0(K,P)-\Gamma_0(0,P)$ is indeed given by the finite equation
\beq\label{GKP}
\Gamma_0(K,P)-\Gamma_0(0,P)&\!=\!&\Lambda_0(K,P)-\Lambda_0(0,P)
\nonumber\\&{}& -\,\frac{1}{2}\int_Q
\left\{\Lambda_0(K,Q)-\Lambda_0(0,Q)\right\}D_{0}^2(Q)\Gamma_0(Q,P).
\eeq
By combining Eqs.~(\ref{eqG}) and (\ref{GKP}), one gets also:
\beq
\Gamma_0(K,P)-\Gamma_0(0,0)&\!=\!&\Lambda_0(K,P)-\Lambda_0(0,0)
\nonumber\\&{}& -\,\frac{1}{2}\int_Q
\left\{\Lambda_0(K,Q)-\Lambda_0(0,Q)\right\}D_{0}^2(Q)\Gamma_0(Q,P)
\nonumber\\&{}& -\,\frac{1}{2}\int_Q
\Gamma_0(0,Q)D_{0}^2(Q)\left\{\Lambda_0(Q,P)-\Lambda_0(Q,0)\right\},
\eeq
an equation similar to that  first obtained by van Hees and Knoll \cite{vanHees:2001ik}.

At this point, we note that the previous analysis departs somewhat from strict perturbation theory: we have
indeed left out at each iteration perturbative contributions which correspond to self-energy insertions in the
propagators of the Bethe-Salpeter equation, or in  the lines of
$\Lambda_0$. Let us then consider the effect of such self-energy insertions. We assume that the added
self-energy  contains all the counterterms needed to make it finite. The question arises then of whether
this insertion brings in new types of
 subdivergences. The answer is clearly negative. Indeed, a new
subdivergence could be obtained only by keeping finite  the momentum in one of the internal line of the added
self-energy. This is equivalent to opening that line, leading to a convergent 6-point function. It follows that
the  structure of subdivergences exhibited 
in the previous analysis is not modified by the replacement of
$D_0$ by another propagator $D$.
 However, contrary to what happens in the case of the 2-loop skeleton where a similar
insertion of a self-energy on the intermediate propagator 
of the Bethe-Salpeter equation only affects the finite part of
$\Gamma$, here  the divergent part is also modified. This is because the 
dominant asymptotic behaviour of $D$ is altered when 
the two-loop self-energy is included in the propagator. This implies
that the numerical values of the counterterms 
obtained at a given iteration of the Bethe-Salpeter equation, such as
$\delta\lambda_0$ in Eq.~(\ref{lambdasunset}), or
$\delta\lambda_1$ in Eq.~(\ref{G0a1}), will 
receive corrections to all orders in the coupling constant coming
from successive insertions of the type we have 
just discussed. We have already met an example of such a correction in the
previous subsection, namely the correction
$\delta\lambda_0^{(1)}$ to $\delta\lambda_0$ (cf. Fig. \ref{2STa}.a). 

It is then useful here  to return to the gap equation in order to see diagrammatically how these various corrections show up
as we proceed through the iterations of this equation.  For illustration, some typical diagrams contributing to 
$\Pi^{(2)}$ at order $\lambda^6$  are displayed in
Fig. ~\ref{2ST}. The  vertex subdivergences are the global divergences
of the 4-point subgraphs enclosed within the dashed line boxes. 
Those in Figs. \ref{2ST}.a--c
correspond to a renormalization of the kernel $\Lambda$ of the Bethe-Salpeter
equation: they can be absorbed in a correction $\delta\lambda_0^{(2)}$ 
of order $\lambda^6$ to $\delta\lambda_0$.
The two boxes in Figs.
\ref{2ST}.d and e correspond to propagator corrections on the first iteration of the Bethe-Salpeter equation:
they contribute to the correction 
$\delta\lambda_1^{(1)}$ of $\delta\lambda_1$ (compare to
Fig. \ref{2STa}.b and c). Finally, the subdivergence in
Fig. ~\ref{2ST}.f corresponds to the second iteration of the Bethe-Salpeter 
equation: the corresponding
divergence is aborbed into $\delta\lambda_2$ (after proper elimination of the subdivergences). 

\vspace{2mm}
\begin{figure}[htb]
\epsfysize=9.cm
\centerline{\epsffile{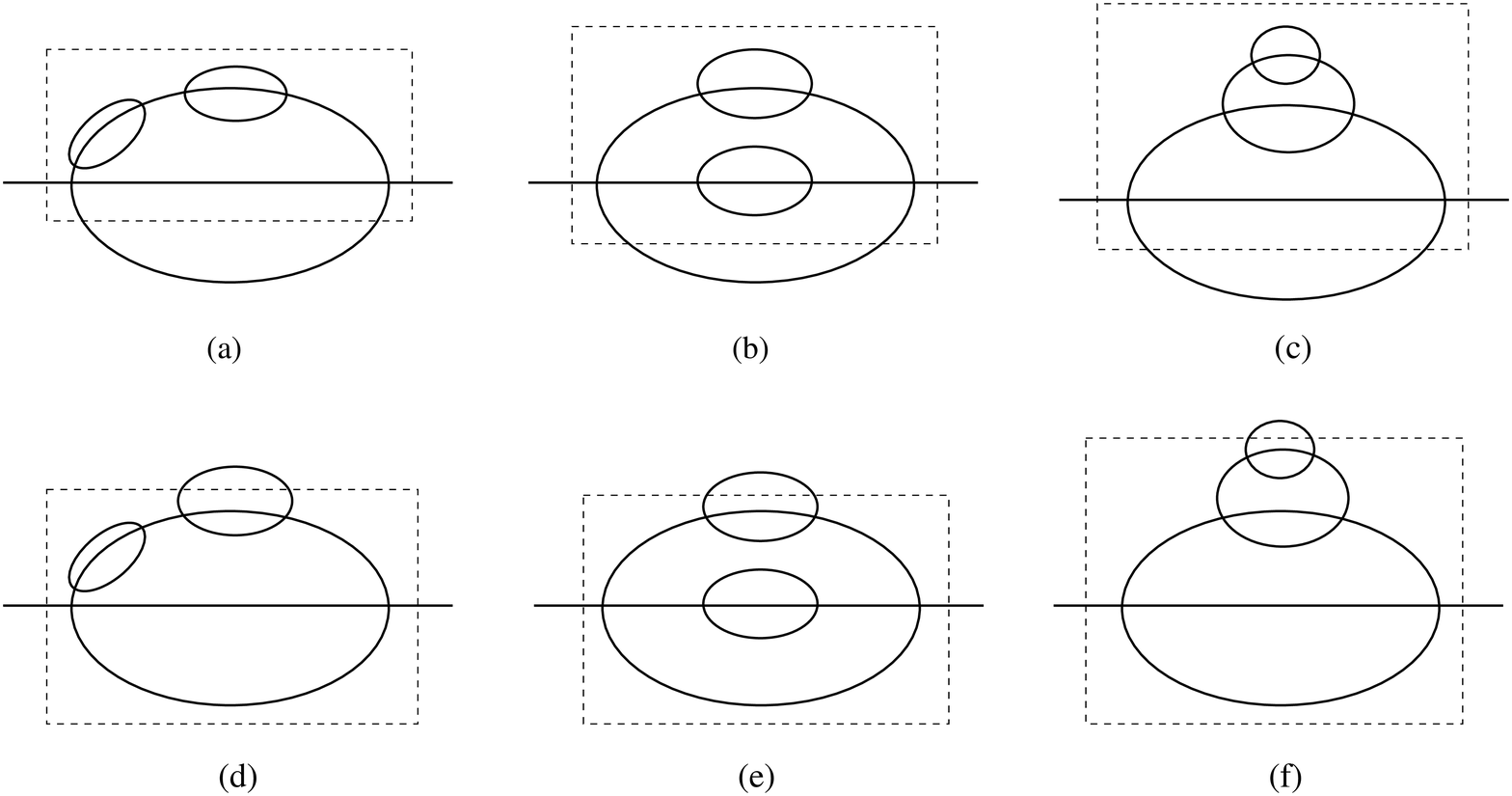}}
\vspace{2mm}
\caption[a]{Diagrams illustrating some of the new subdivergences
which occur at order $\lambda^6$. The divergent subgraphs are enclosed in the dashed
line boxes.
\label{2ST} }
\end{figure}

We  turn now to the mass counterterm $\delta m^2$.
As we have seen, this can be deduced from the 
equation obtained by differentiating the gap equation with respect to $m^2$. 
We can rewrite Eq.~(\ref{m2dPidm2bR}) as follows:
\beq\label{m2dPidm2c}
\int_Q\left[\delta(K-Q)+\frac{1}{2}\Lambda(K,Q)D^2(Q)\right]\frac{\del \Pi(Q)}{\del
m^2}=-\frac{1}{2}\int_Q \Lambda(K,Q)D^2(Q)+\frac{\delta m^2}{m^2}.
\eeq
This equation has a simple diagrammatic interpretation similar to 
that given in  Fig.~\ref{fig:dPidm2}, with now
$\Lambda$ replacing the coupling constant of Fig.~\ref{fig:dPidm2}. The diagrams contributing
to 
$\del\Pi/\del m^2$ that are generated by this equation have vertex and mass
subdivergences, analogous to those of the self-energy itself. It is however easy to verify that
the vertex subdivergences are eliminated by the same counterterms as those which make
finite the Bethe-Salpeter equation. This is most readily seen on the following form 
of the equation:
\beq\label{eqfordeltam2}
\frac{\del \Pi(P)}{\del m^2}=-\frac{1}{2}\int_Q \Gamma(P,Q)D^2(Q)+\frac{\delta m^2}{m^2}
\int_K\left[\delta(P\!-\!K)-\frac{1}{2}\Gamma(P,K)D^2(K)\right],
\eeq
obtained from Eq.~(\ref{m2dPidm2c}) by using Eqs.~(\ref{G0cx})--(\ref{G0cy}).
This expression of $\del \Pi/\del m^2$ in terms of $\Gamma$ contains no vertex
subdivergences. The remaining
divergences can be absorbed in the mass counterterm, determined then by requiring that
$\del\Pi/\del m^2$ be finite. 
We shall illustrate this cancellation of subdivergences, by looking at a few
iterations of Eq.~(\ref{eqfordeltam2}).
For the sake of simplicity, we use in all expressions the free propagator
$D_0$, rather than the full propagator $D$ (as we have already mentioned, this does not change the structure
of the counterterms, only their numerical values; in particular the way the various subdivergences cancel is
independent of this choice). In zeroth order we have
\beq\label{deltam2sunset0}
\frac{\delta m^2_0}{m^2}=\left[\frac{1}{2}\int_Q \tilde\Lambda_0(P,Q)D_0^2(Q)\right]_{\mbox{\tiny div}},
\eeq
where the symbol $\left[\cdots\right]_{\mbox{\tiny div}}$ isolates the divergent part of the quantities inside the
brakets. One can verify that Eq.~(\ref{deltam2sunset0}) leads to the expression
(\ref{deltam2sunsetpt}) of the mass counterterm obtained in perturbation theory.  The first
iteration gives
\beq
\frac{\delta
m^2_1}{m^2}\!=\!-\frac{1}{2}\left[\int_Q \Gamma_0^{(1)}(P,Q)D^2(Q)\right]_{\mbox{\tiny div}}\!\!+\!\frac{\delta
m^2_0}{m^2}
\int_K\left(\!\delta(P\!-\!K)\!-\!\frac{1}{2}\tilde\Lambda_0(P,K)D_0^2(K)\!\right)\!,
\eeq
which after using Eq.~(\ref{G0a1}) leads to 
\beq\label{deltam2sunset1}
\frac{\delta
m^2_1}{m^2}&=&\frac{1}{2}\left[\int_Q \left(\delta\lambda_1+\frac{1}{2}\int_K
\tilde\Lambda_0(P,K)D^2_0(K)\tilde\Lambda_0(K,Q)\right) D_0^2(Q)\right]_{\mbox{\tiny div}}
\nonumber\\ &+&\frac{\delta
m^2_0}{m^2}\int_K \frac{1}{2}\tilde\Lambda_0(P,K)D_0^2(K).
\eeq
In contrast to the case of the 2-loop skeleton, 
where the multiple integrals factorize, here one cannot 
obtain an explicit expression. One can however easily 
recognize the role of counterterms determined at a previous iteration (here
$\delta m_0^2$), or those coming from the Bethe-Salpeter equation 
(here $\delta\lambda_1$), in the
cancellation of subdivergences. Observe in particular that the product of
$\delta m_0^2$ by the finite part of the integral in the second line, which
is $P$--dependent, cancels with a corresponding term in the first line
which arises by keeping $K$ fixed and taking $Q$ to infinity.

\subsection{The determination of $\delta Z$ and the asymptotic 
form of $\Pi$\label{sec:deltaZ}}

In the discussion of the two-loop approximation in Sect. II, we have been
able to separate the calculation of the vertex and mass counterterms $\delta\lambda$ and
$\delta m^2$ from the resolution of the gap equation for the propagator.
This has been possible since the divergences appearing in the Bethe-Salpeter equation for $\Gamma$,
and also in Eq.~(\ref{m2dPidm2a1}) for ${\del \Pi}/{\del m^2}$, depend only upon
the leading behaviour of the propagator at large momenta, and this is
not modified by the resummation performed by the one-loop gap equation (\ref{Pitad}).
Therefore, in order to compute the counterterms $\delta\lambda$ and
$\delta m^2$, it was enough to consider these equations with the full propagator
replaced by the free one, $D_0(P)=(P^2+m^2)^{-1}$ (where the mass $m$ did not play
any role except that of an infrared regulator).

But in the present, three-loop approximation, such simplifications no longer hold.
Indeed, as revealed by the analysis of
 the first orders of perturbation theory, the insertion of the
self-energy does change
the asymptotic behavior of the propagator, so that this behavior keeps changing
as one proceeds through the iterations.
Thus, if one were to compute the propagator and the various counterterms
through iterations, both the gap equation and the equations for $\delta\lambda$ and
$\delta m^2$ would have to be solved simultaneously. Still, as we explain now, it is
nevertheless possible to disentangle 
the calculation of the counterterms from that of the full solution
to the gap equation. The key observation is that the counterterms are sensitive only
to the {\it leading} asymptotic behavior of the propagator. This is clear in 
the previous perturbative analysis, and
in the general case it results from the fact that
Eqs.~(\ref{BSdef}) and (\ref{eqfordeltam2}) involve only
logarithmic divergences. As for $\delta Z$, it is determined by the global
divergence of the self-energy diagrams obtained when all the internal loop momenta
are simultaneously sent to infinity (after all subdivergences
have been removed): this divergence is proportional to $K^2$,
and is independent of the mass, or any other subleading contribution
to the propagator at large momenta.

Now, according to
Weinberg's theorem \cite{Weinberg,Collins:xc},  
one expects that for very large momenta $K\gg m$, 
$\Pi(K)/K^2$ grows, at most, as a (slowly varying) function of $\ln K^2$.
It is useful to introduce the special notation $\Pi_2(K)$ for the leading
asymptotic contribution:  by assumption $(\Pi-\Pi_2)/K^2 \to 0$ as $K^2 \to \infty$. To compute
$\Pi_2(K)$, we need the limit of Eq.~(\ref{eq:gapA}) for very high $K^2$ (and zero
temperature). Since the mass plays no role in this limit, one could set $m=0$,
and we shall indeed do that eventually. However in practice, it is sometimes useful to keep
a non-zero mass to avoid infrared complications, so let us consider first the
case $m\ne 0$. The corresponding gap equation involves all types of
divergences, but, for the purpose of computing $\Pi_2$, it can here be renormalized
by simple subtractions. That is, $\Pi_2(K)$ can be obtained from the large $K$ 
behavior of the solution $\Pi$ to the following, auxiliary, gap equation:
\beq\label{Pi2sunset0}
\Pi(K)\,=\,S(K^2) - S(K^2=-m^2) - (K^2+m^2)\frac{\del S}{\del K^2}\Big|_{K^2=-m^2} .
\eeq
where $S(K)$ denotes the r.h.s. of Eq.~(\ref{eq:gapA}).
This equation has been used in Ref. \cite{VanHees:2001pf}
to compute the self-energy in the vacuum. It is finite indeed,
as it can be checked by taking two derivatives
with respect to $K^2$ within the integrand, and noticing that the result is finite.
It should be emphasized that this procedure holds because, in the present approximation,
 the vertex counterterm $\delta\lambda$ enters only a local self-energy 
diagram (the tadpole in the r.h.s. of Eq.~(\ref{eq:gapA})), and thus it can be
simulated by a local, mass, subtraction. In any case, Eq.~(\ref{Pi2sunset0}) is
correct only for the calculation of $\Pi_2$, and not for
the subleading contribution $\Pi_0\equiv \Pi-\Pi_2$.

In what follows we shall compute $\Pi_2$ from the gap
equation with $m=0$, using dimensional regularization. Then, both the mass 
divergences  and the vertex subdivergences automatically cancel out, and the remaining field 
divergences can be eliminated by the counterterm $\delta Z K^2$.
The equation determining $\Pi_2$ reads then
\beq\label{Pi2sunset}
\Pi_2(K)\,=\,S(K) \, +\,K^2\delta Z
\eeq
where this time
$S(K)$ denotes the r.h.s. of Eq.~(\ref{eq:gapA})
evaluated with propagator $D_2^{-1}(K)\equiv K^2 +\Pi_2(K)$.
 Clearly, the finite  form of the equation,
and thus also the ensuing solution for $\Pi_2$, will depend upon our choice
of a subtraction scheme. For instance, one can choose $\delta Z = - 
(\del S/\del K^2)|_{K^2=-\mu^2}$, in which case the asymptotic behaviour of the
propagator will depend upon the subtraction scale $\mu$.
Since, as we have seen before,
the counterterms $\delta\lambda$ and $\delta m^2$ (including their divergent parts)
are sensitive to the asymptotic behavior of the propagator, it follows that these 
counterterms will end up being dependent upon the renormalization scheme for $\delta Z$.

We can now summarize the analysis of the last two subsections as follows:
{\it i}) The values of the counterterms are determined by the {\it dominant} piece of the
self-energy at large $K$. {\it ii}) They can therefore be determined from the solution of an auxiliary
gap equation which coincides at  asymptotically large $K$ with the original gap equation. {\it iii}) The
coupling renormalization involves two types of counterterms: those which renormalize the kernel
$\Lambda$ of the Bethe-Salpeter equation,  and those which render finite the iterations of the
Bethe-Salpeter equation. The counterterms of the first kind can be summed up as the counterterm
$\delta\lambda_0$. The counterterms of the second kind can be computed in terms of the solution
to the Bethe-Salpeter equation. {\it iv}) The mass 
counterterms are determined by the equation for $\del \Pi/\del m^2$
where the
vertex subdivergences are eliminated by the counterterms which render finite the Bethe-Salpeter
equation.

\subsection{The gap equation: algebraic renormalization}

The verification that, with the counterterms determined as indicated above,  the gap equation is indeed finite, goes through
similar steps as  in  subsection \ref{sec:tadpolealgebraic}. The main complication here comes from the need
to deal with the modification of the asymptotic part of the propagator resulting from the insertion  of the momentum
dependent self-energy. We shall handle this by considering an auxiliary gap equation, as indicated in the
previous subsection.

We write the  full propagator as
\beq
D(K)=\frac{1}{K^2+m^2+\Pi(K)}
\eeq
where $\Pi(K)$ is the renormalized self-energy. Then we set 
\beq\label{PiPI0PI2}
\Pi=\Pi_0+\Pi_2 
\eeq
where $\Pi_2$ is the solution of an auxiliary gap equation which coincides asymptotically with
the original gap equation. As indicated in the previous subsection, $\Pi_2$ may be 
obtained, for instance,  as the solution of the gap equation with $m=0$, that is:
\beq\label{auxiliarygap}
\Pi_2(K)\!=\!-\frac{\lambda^2}{6}\int_P\int_Q D_{20}(P)D_{20}(Q)D_{20}(K+P+Q)+K^2\delta Z,
\eeq
where $D_{20}(P)^{-1}\equiv P^2+\Pi_2(P)$. Note that, thus defined, $\Pi_2(0)=0$. 
 We denote by
$\Pi_0(K)$ the difference between the true solution, $\Pi$, and $\Pi_2$. It follows from
our  assumptions that:
\beq
\lim_{K\to \infty}\frac{\Pi_0(K)}{K^2}=0,
\eeq
so that  
$\Pi_0(K)$ grows at most logarithmically at large $K$. There is of course some
arbitrariness in the decomposition (\ref{PiPI0PI2}), related to the specific choice  one makes for $\Pi_2$.
The important point is that the asymptotic behaviour of
$\Pi_2(K)$ at large $K$ coincides with that of $\Pi(K)$. Indeed, from the previous discussions,
we know that the counterterms depend  only on $\Pi_2$ and are insensitive to the precise separation
between $\Pi_2$ and $\Pi_0$.

The propagator
\beq
D_{2}(K)\equiv (K^2+m^2+\Pi_2)^{-1} 
\eeq
 will play the same role here as $D_{0}(K)$ in the previous section,
i.e., we shall expand with respect to it.  We include $m^2$
within $D_{2}$ in order to avoid possible infrared problems in this expansion (the massless case is
discussed in Sec.~\ref{sec:masslesssunset}).    We  call
$\Gamma_2$ the solution of the Bethe-Salpeter equation where the intermediate propagators are $D_2$
propagators and similarly the kernel
$\Lambda_2$ is evaluated with $D_2$ propagators. We shall assume that the counterterms
$\delta\lambda$ and $\delta m^2$ are determined from this equation and the corresponding one for
$\del\Pi/\del m^2$. That is, the solution of the auxiliary gap equation provides a finite
$\Pi_2$ and well defined values for the counterterms $\delta Z$, $\delta \lambda$ and
$\delta m^2$. We verify now that these counterterms render the 
initial gap equation finite. The strategy will be to use the auxiliary gap equation
to obtain $D_2$ and then determine the counterterms. These counterterms will
then be used in order to eliminate the divergences of the gap equation, leading
to a finite equation for $\Pi_0$.

It is useful first to introduce a special notation for the self-energy calculated 
with the propagator $D_2$. We set $\tilde\Pi_2(K)=\Pi[D_2]$, where $\Pi[D]$ denotes the right hand side
of  Eq.~(\ref{eq:gapA}); that is: 
\begin{equation}\label{eq:gapA2}
\tilde\Pi_2(K)\!=\!-\frac{\lambda^2}{6}\int_P\int_Q D_2(P)D_2(Q)D_2(K+P+Q)+
\,\frac{\delta \lambda}{2}\int_P D_2(P)+\delta
m^2+K^2\delta Z.
\end{equation}
Note that if the counterterms $\delta\lambda$ and $\delta m^2$ in this equation are limited to
their leading order values, $\delta\lambda_0$ and $\delta m^2_0$ 
respectively 
(calculated with propagator $D_2$), then $\tilde\Pi_2(K)$ is finite.

We now expand the propagator $D$ around $D_2$
\beq\label{expand DD2}
D=D_{2}+\delta D\qquad \delta D=D_{2}[-\Pi_0]D_{2}
+D_{r}
\eeq
 where we have singled out in 
$\delta D$ the term linear in  $\Pi_0$. The usefulness of this expansion is that each extra power of
$\Pi_0$ decreases the degree of divergence of the diagram in which the corresponding propagator is
inserted. This property will allow us to isolate the vertex subdivergences in very much the same way as
in  the two-loop example. In analogy to what we did in
Sect.~\ref{sec:tadpolealgebraic} we define
 \beq\label{tildePi0def}
  \tilde{\Pi}_0(K)=\frac{1}{2}\int_P \Lambda_2(K,P)\delta
D(P).
 \eeq

Now we note that the gap equation can be written as $\Pi=\Pi[D]$ where $D$ is expanded as in
Eq.~(\ref{expand DD2}), and may thus be considered a function of $\Pi_0$. Using the definitions
(\ref{eq:gapA2}) and (\ref{tildePi0def}) above, we can put the gap equation in the form
  \beq\label{tildegapeqn}
\Pi=\tilde
\Pi_2+\tilde\Pi_0+\tilde\Pi_r
  \eeq
   where $\Pi$ in the left hand side is to be understood as $\Pi_2+\Pi_0$ (so that the equation is
efffectively an equation for $\Pi_0$). The quantity 
$\tilde\Pi_r(K)$  is not only finite, 
but decreases  as $1/K^2$ at large $K$. This follows from Weinberg's
theorem and the fact that
$\tilde\Pi_r$ contains at least two $\Pi_0$ insertions and is therefore of degree of
divergence $-2$.

Next we note that 
\beq
\tilde{\Pi}_0(K)-\tilde{\Pi}_0(0)=\frac{1}{2}\int_P\left[\Lambda_2(K,P)-\Lambda_2(0,P)\right]\delta
D(P)
\eeq
 is finite (cf. after Eq.~(\ref{eqG})).  Since both $\tilde\Pi_0(K)- \tilde\Pi_0(0)$ and
$\tilde\Pi_2(K)- \tilde\Pi_2(0)$ are finite, so is therefore $\Pi(K)-\Pi(0)$. It follows
that the ultraviolet divergences are entirely contained in $\Pi(0)=\tilde
\Pi_2(0)+\tilde\Pi_0(0)+\tilde\Pi_r(0)$ where both $\tilde
\Pi_2(0)$ and $\tilde\Pi_0(0)$ are a priori divergent.

For $\tilde\Pi_0(0)$ we proceed as in Sect.~\ref{sec:tadpolealgebraic}  and express
$\Lambda_2(0,P)$ in the defining equation for $\tilde
\Pi_0(0)$ (Eq.~(\ref{tildePi0def}) with $K=0$) in terms of $\Gamma_2(0,P)$
(the solution to the Bethe-Salpeter equation with propagator $D_2$). We get:
\beq\label{tildePi0sansPi}
\tilde{\Pi}_0(0)&\!\!=\!\!&\frac{1}{2}\int_P \Lambda_2(0,P)
\delta D(P)\nonumber\\ 
&\!\!=\!\!&\frac{1}{2}\int_P\left\{  
\Gamma_2(0,P)+\frac{1}{2}\int_Q \Gamma_2(0,Q) 
D_{ 2}^2(Q) \Lambda_2(Q,P) 
\right\}\delta D(P)\nonumber\\
&\!\!=\!\!&\frac{1}{2}\int_P\Gamma_2(0,P)\left[\delta D(P)+
D_{ 2}^2(P)\tilde\Pi_0(P)\right]
\nonumber\\
&\!\!=\!\!&\frac{1}{2}\int_P\Gamma_2(0,P)\left[D_r(P)+
D_{ 2}^2(P)(\tilde\Pi_0(P)-\Pi_0(P))\right]
\eeq
For $\Pi$ solution of the gap equation, we can use Eq.~(\ref{tildegapeqn}) to replace 
$\tilde{\Pi}_0$ by $\Pi-\tilde\Pi_2-\tilde\Pi_r$ in the r.h.s. of this equation, and
verify that the divergent terms linear in $\Pi_0$
($=\Pi-\Pi_2$)  cancel, as anticipated. We are left with:
\beq\label{eq:gap_ren}
\tilde\Pi_0(0) &\!\! =\!\! &\frac{1}{2}\int_P
\Gamma_2(0,P)\left\{
D_{r}(P)-\tilde{\Pi}_r(P)D_{2}^2(P)\right\} \nonumber\\
   & \!\!+\!\!&\frac{1}{2}
    \int_P \Gamma_2(0,P)\left\{
{\Pi}_2(P)-\tilde{\Pi}_{2}(P)\right\}D_{2}^2(P).
\eeq
The first line is finite, as it can be easily checked by using 
$D_{r}(P)\sim 1/P^6$, $\tilde{\Pi}_r(P)\sim 1/P^2$, and the fact
that $\Gamma_2(0,P)$ behaves like $\ln P^2$.

 The remaining divergences  in  $\Pi(0)$ are
therefore those of the following expression (we add $\tilde\Pi_2(0)=0$):
\begin{equation}\label{eq:remaining}
\frac{1}{2}\int_P\Gamma_2(0,P)\left\{\Pi_2(P)\!-\!\tilde{\Pi}_2(P)\right\}D_{2}^2(P)+\tilde{\Pi}_{2}(0).
\end{equation}
We  verify now that the counterterm $\delta m^2$ absorbs the potential divergence. 
To do so, observe first that this expression vanishes in the limit $m\to 0$ (then $\tilde
\Pi_2(P)=\Pi_2(P)$ and $\Pi_2(0)=0$). Let us then expand
$\Pi_2-\tilde{\Pi}_2=\Pi[D_{20}]-\Pi[D_{2}]$ around $D_{2}$ by using :
\begin{equation}\label{expD20}
D_{20}-D_{2}=\frac{1}{P^2+\Pi_2(P)}-\frac{1}{P^2+m^2+\Pi_2(P)}=m^2D^2_{2}(P)+D_{s}(P),
\end{equation}
where $D_{s}(P)\sim 1/P^6$ at large $P$.
We get
\begin{equation}\label{Pi2moinstildePi2}
\Pi_2(P)\!-\!\tilde{\Pi}_2(P)=-\delta m^2+\frac{1}{2}m^2\int_Q
\Lambda_2(P,Q)\,D^2_{2}(Q)+\Pi_{s}(P),
\end{equation}
where $\Pi_{s}(P)$ is finite and goes as $1/P^2$ at large $P$. 
Inserting  this in Eq.~(\ref{eq:remaining}), and keeping only  the terms which are not manifestly finite, we obtain:
\begin{eqnarray}\label{Pi2moinstildePi2b}
& &
\frac{m^2}{2}\int_Q\left\{\frac{1}{2}\int_P\Gamma_2(0,P)\,D^2_{2}(P)\Lambda_2(P,Q)-\Lambda_2(0,Q)\right\}D_{2}^2(Q)
\nonumber\\
& & +\delta m^2\left\{1-\frac{1}{2}\int_P\Gamma_2(0,P)D_{2}^2(P)\right\}\nonumber\\
\end{eqnarray}
One recognizes in this expression the right hand side of Eq.~(\ref{eqfordeltam2}) which specifies the value of the mass
counterterm. It is therefore finite.

\subsection{Finite temperature}
\label{sec:finiteT}

It is straightforward to extend the diagrammatic analysis presented earlier in this section to the case of finite
temperature. Doing so,  one  easily verifies that, iteration by iteration, all contributions involving the temperature are
finite. Potential divergences coming from 
counterterms multiplying finite temperature contributions cancel out  when
subdivergences are eliminated. Furthermore, since the counterterms depend only on the asymptotic part of the propagator,
which is unaffected by finite temperature effects, 
they are manifestly independent of the temperature.
(The fact that $\Pi_2(P)$ is not modified by the temperature follows
from Weinberg's theorem \cite{Weinberg,Collins:xc}, 
which ensures that the asymptotic behaviour 
$\sim P^2 F(\ln P^2)$ of the self-energy comes from vacuum diagrams
in which the internal momenta in all the internal lines are large
and proportional to $P$.)

But although this diagrammatic analysis reveals that no difficulty of principle should be 
expected, in actual 
calculations  the temperature dependent contributions may become mixed in a subtle way with 
vacuum, divergent
contributions, and some work may be required to disentangle them.  To be specific, 
we consider here the calculation in the
imaginary time formalism. It is possible to organize the summation over the Matsubara frequencies 
in such a way as to
separate the explicit thermal dependence in terms of integrals with 0, 1 and 2 thermal factors:
\begin{equation}\label{Pi012}
\Pi(K)=\Pi_{\mbox{\tiny 0n}}(K)+\Pi_{\mbox{\tiny 1n}}(K)+\Pi_{\mbox{\tiny 2n}}(K).
\end{equation}
The first contribution, $\Pi_{\mbox{\tiny 0n}}(K)$ 
is exactly the same functional of $D$ as at zero temperature,
Eq.~(\ref{eq:gapA}).  Since $D$ depends
 on the temperature (because $\Pi$ does),  $\Pi_{\mbox{\tiny 0n}}(K)$ depends
implicitly on the temperature. 
The term with one statistical factor, $\Pi_{\mbox{\tiny 1n}}(K)$, can be written as: 
\beq
\Pi_{\mbox{\tiny 1n}}(K)=\frac{1}{2}\int_{\tilde{P}}\sigma(\tilde{P})\Lambda (K,\tilde{P})
\eeq
where $ \Lambda $ is the same functional 
of $D$, Eq.~(\ref{lambdasunset0}), as at zero temperature and
 depends only implicitly on the temperature through $D$. We have set 
$\sigma(\tilde{P})=\rho(p_0,p)n_{|p_0|}\varepsilon_{p_0}$, 
where $\rho(p_0,p)$ is the spectral function for the propagator
$D$:
\beq
D(\omega,p)=\int_{-\infty}^\infty \frac{{\rm d}p_0}{2\pi}\,\frac{\rho(p_0,p)}{p_0-\omega},
\eeq
  $n_p=1/({\rm e}^{p/T}-1)$, and $\varepsilon_{p_0}=\pm 1$ depending on the sign of $p_0$. Finally, we have introduced the
notation $\tilde{P}=(p_0,p)$, to be distinguished from 
$P=(ip_0,p)$. The 4-point functions such as 
$\tilde\Lambda_2(K,\tilde{P})$ are analytically continued in the complex plane from the complex value $ip_0$ 
 where they are well defined (note that since $\Lambda(K,P)$ is a function of $K+P$, it remains well
defined when $P$ is continued to $\tilde P$).
Finally the term with two statistical factors, 
$\Pi_{\mbox{\tiny 2n}}(K)$, can be written as: 
\beq\label{eq:tildeP2}
\Pi_{\mbox{\tiny 2n}}(K)=\frac{\lambda^2}{2}\int_{\tilde{P}}
\int_{\tilde{Q}}\sigma(\tilde{P})\sigma(\tilde{Q})D(K-\tilde{P}-\tilde{Q}).
\eeq

As in the previous subsection, we write $\Pi=\Pi_2+\Pi_0$, with 
$\Pi_2$ determined by the zero-temperature auxiliary gap equation (\ref{auxiliarygap}), and $\Pi_0(K)$ is at most
logarithmic and carries the whole dependence upon $T$. For $\Pi_{\mbox{\tiny 0n}}(K)$ we set again:
\begin{equation}\label{Pion20r}
\Pi_{\mbox{\tiny 0n}}(K)=\tilde{\Pi}_2(K)+\tilde{\Pi}_0(K)+\tilde\Pi_r(K),
\end{equation}
where $\tilde{\Pi}_2$, $\tilde{\Pi}_0$, and $\tilde\Pi_r$ are given by the same Euclidean integrals as in the previous
subsection, but contain implicit temperature dependence through the propagator $D$, except for $\tilde \Pi_2$ which
depends only on $D_2$ and remains therefore temperature independent. It is also convenient to write
\begin{equation}
\Pi_{\mbox{\tiny 1n}}(K)=\tilde{\Pi}_1(K)+\tilde{\Pi}_3(K),
\end{equation}
where
\beq\label{eq:tildeP1}
\tilde{\Pi}_1(K)=\frac{1}{2}\int_{\tilde{P}}\sigma(\tilde{P}) \Lambda_2(K,\tilde{P})
\eeq
is at most logarithmic at large $K$, and can  therefore be considered as a modification of $\tilde\Pi_0$. As for
$\tilde{\Pi}_3$, it decreases faster than a logarithm and it will be considered as a modification of $\tilde \Pi_r$. The
same remark applies to 
$\Pi_{\mbox{\tiny 2n}}(K)$ that we shall denote by $\tilde{\Pi}_4(K)$.

 The final expression for the gap equation takes then a form
similar to that at zero
temperature, and may be written as
\beq
\Pi=\tilde\Pi_2+(\tilde\Pi_0+\tilde\Pi_1)+(\tilde\Pi_r+\tilde\Pi_3+\tilde\Pi_4)=\tilde\Pi_2+\tilde\Pi_0'+\tilde\Pi_r',
\eeq
with $\tilde\Pi'_0=\tilde\Pi_0+\tilde\Pi_1$ and  $\tilde\Pi'_r=\tilde\Pi_r+\tilde\Pi_3+\tilde\Pi_4$.  Applying the
 strategy of the previous subsection to eliminate the vertex subdivergences in $\tilde\Pi'_0(0)$, one gets  easily:
\begin{eqnarray}\label{rengapT}
\tilde\Pi'_0(0) & = & \frac{1}{2}\int_{P}D_r(P)\Gamma_2(0,P)+\frac{1}{2}\int_{\tilde{P}}\sigma(\tilde{P})\Gamma_2(0,\tilde{P})\nonumber\\
& - & \frac{1}{2}\int_P \Gamma_2(0,P)D_2^2(P)\left\{\Pi_2(P)-\tilde\Pi_2(P)-\tilde\Pi'_r(P)\right\}.\nonumber\\
\end{eqnarray} 
This equation generalize Eq.~(\ref{eq:gap_ren}). Its first line is finite. The divergences in the second line  are
temperature independent. When combined with those of $\tilde\Pi_2(0)$, they become identical to
those of  the zero temperature case, and are   absorbed in 
$\delta m^2$.

 \subsection{Massless case\label{sec:masslesssunset}}  

The massless case is interesting only at non-zero temperature. (Indeed, if $T=0$, then the gap
equation with $m=0$ is simply the auxiliary gap equation that we have introduced before,
cf. Eq.~(\ref{auxiliarygap}), and whose solution is $\Pi_2$.) Consider then the finite--temperature
case, where $\Pi$ admits the decomposition shown in Eq.~(\ref{Pi012}). When $m=0$, the
subsequent manipulations of the term with zero thermal factors, $\Pi_{\mbox{\tiny 0n}}$,
 deserve a special
treatement: Indeed, if one proceeds as before, i.e., by developing $D$ around $D_2$ as in
Eq.~(\ref{expand DD2}), one meets infrared problems:   since $D_2$ is now
 a massless propagator (this is the
same as $D_{20}$ introduced before), the Bethe-Salpeter equation for the 4-point function 
$\Gamma_2$ is infrared divergent, and so are also $\tilde\Pi_r$ and $D_r$ used in (\ref{eq:gap_ren}).
One can however follow the same strategy as in Sec.~\ref{sec:masslesstadpole} and exploit the fact that the finite temperature effects generate a natural infrared cutoff. Since
the ultraviolet divergences are ultimately sensitive only to 
the asymptotic behaviour of the propagator,  they will not be affected by the
replacement of $D_2$
by $D$ at appropiate places in the previous expansions. Specifically, we shall
 exploit the fact that $\Pi_2$ is the solution to the massless gap equation at $T=0$,
namely (cf. Eq.~(\ref{auxiliarygap})) : $\Pi_2=\Pi_{\mbox{\tiny 0n}}[D_2]$. Then, after rewriting the
full propagator as in Eq.~(\ref{expansionD2}):
\begin{equation}
D_{2}(P)=D(P)+D(P) \Pi_0(P) D(P)-D'_r(P)\,\equiv\,D -\delta D'\,,
\end{equation}
one can develop $\Pi_{\mbox{\tiny 0n}}[D_2]$ around $\Pi_{\mbox{\tiny 0n}}[D]$ 
without encountering infrared divergences:
\beq
\Pi_{\mbox{\tiny 0n}}[D_2]\,=\,\Pi_{\mbox{\tiny 0n}}[D]\,-\,
\frac{1}{2}\int_P \Lambda(K,P)\delta D'(P)\,-\,\tilde\Pi_r^\prime(K),
\eeq
or, equivalently (compare to Eq.~(\ref{Pion20r})):
\begin{equation}\label{Pion20rnew}
\Pi_{\mbox{\tiny 0n}}(K)\,=\,{\Pi}_2(K)\,+\,\frac{1}{2}\int_P \Lambda(K,P)\delta D'(P)
\,+\,\tilde\Pi_r^\prime(K),
\end{equation}
where $\Pi_{\mbox{\tiny 0n}}\equiv \Pi_{\mbox{\tiny 0n}}[D]$.
Note that, as compared to Eq.~(\ref{tildePi0def}), the development above involves
the kernel $\Lambda$ built with the full propagator, and similarly for $\delta D'$.
It is then
 possible to use the complete Bethe-Salpeter equation (with propagator $D$) in order
to replace $\Lambda$ in terms of the infrared finite 4-point function
$\Gamma$.

The subsequent manipulations proceed as before, in Sect. \ref{sec:finiteT}, except for
the replacements $\Gamma_2\to \Gamma$ and $\delta D\to \delta D'$. Since $\Pi_2\equiv \tilde\Pi_2$
when $m=0$, this finally leads to the following, manifestly finite, expression for
$\tilde\Pi_0^\prime$ (compare to Eq.~(\ref{rengapT})):
\begin{eqnarray}\label{rengap0T}
\tilde\Pi_0^\prime(K) 
\,= \,\frac{1}{2}\int_{P}\Gamma(K,P)\Big[D_r^\prime(P) - \tilde\Pi_r^\prime(P)D^2(P)
\Big]\,+\,\frac{1}{2}\int_{\tilde{P}} \sigma(\tilde{P})\Gamma(K,\tilde{P})\,.
\end{eqnarray} 
Since, however, the Bethe-Salpeter equation for $\Gamma$ involves the full propagator $D$,
and thus is sensitive to the complete self-energy $\Pi={\Pi}_2 + \tilde\Pi_0 + \tilde\Pi_r^\prime$,
it follows that, unlike in the massive case, the Bethe-Salpeter and the (full) gap equation
are now coupled, and must be solved simultaneously.

    \section{Generalization to higher-loop order}  

In order to generalize the previous considerations to higher-loop order skeletons, we note first that many
of the results that have been obtained in the previous section are independent of the precise
structure of the skeleton considered. In particular, the equations which determine the vertex and mass
counterterms, i.e. the Bethe-Salpeter equation (\ref{RenormBS}) and the equation
(\ref{eqfordeltam2}) for
$\del\Pi/\del m^2$,  have been used in the previous section in their general forms, without
further specifying the diagrammatic structure of $\Phi$, $\Pi$, or $\Lambda$.
Similarly,  the manipulations performed on these equations, like the expansion
of the gap equation in Eq.~(\ref{tildegapeqn}), or the expression
(\ref{tildePi0def}) for $\tilde{\Pi}_0(K)$, are generic as well, and would
hold for any choice of the skeletons in $\Phi[D]$. There are new features however 
that need to be discussed. They are already present  in the 4-loop approximation to $\Phi[D]$, and we shall 
illustrate our discussion with this example. The corresponding diagrams are displayed in Fig.~\ref{4loopPHI}.

\vspace{2mm}
\begin{figure}[htb]
\epsfysize=5.cm
\centerline{\epsffile{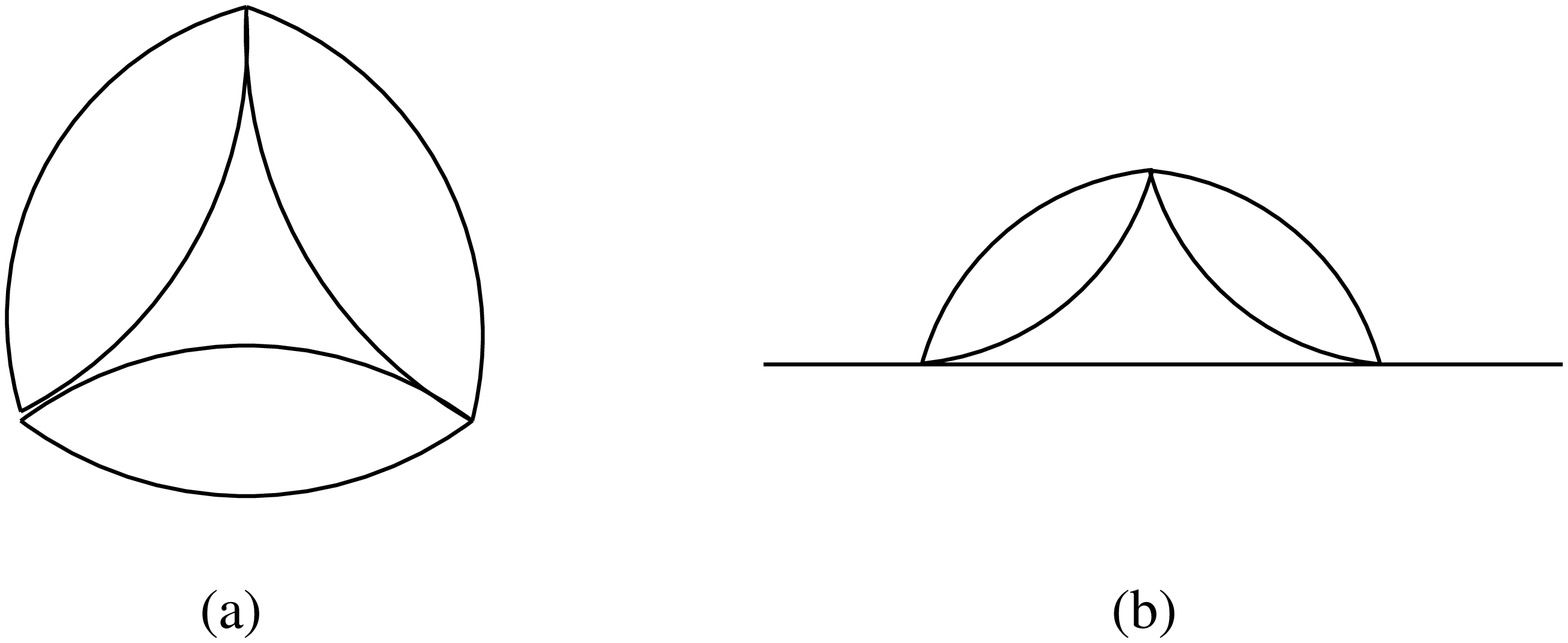}}
\vspace{2mm}
\caption[a]{The 4--loop contribution to $\Phi$ (a) and the
corresponding 3--loop contribution to $\Pi$ (b).
\label{4loopPHI} }
\end{figure}

 As we have seen  in the discussion of the 3--loop 
approximation in Sect. IV, skeleton diagrams with different number of loops 
get mixed under renormalization: the two-loop skeleton is indeed needed as a counterterm to eliminate vertex
subdivergences of the three-loop skeleton. We shall call 
{\it primary skeletons} the skeletons which are selected
in the $\Phi$-derivable approximation that one considers, 
and {\it counterterm skeletons} those skeletons which
carry the counterterms needed to remove the vertex 
subdivergences of the primary skeletons. To be specific, let us
order the skeleton diagams according to their  number of  loops, 
and consider the approximation in which the
primary skeletons are $(N+1)$-loop diagrams (so that the corresponding diagrams for $\Pi$ involve
$N$ loops). The removal of divergences in the gap equation for $\Pi$ proceeds in three steps:

{\it i\,}) In the first step, one determines the 4-point function $\Lambda$
from  $\Pi$ by using Eq.~(\ref{Lambdadef}), and one constructs the vertex counterterms
which render  $\Lambda$ finite. Since 
obtained by opening one line in $\Pi$, the 4-point function $\Lambda$
is given by primary diagrams with $N-1$ loops. In general, these diagrams will involve 
vertex subdivergences. The corresponding counterterms 
are constructed as in ordinary perturbation theory, with however the propagator $D$
associated to the lines of the diagrams; that is, they are the
divergent parts of the loop integrals associated to the respective (sub)graphs
evaluated with  propagator $D$.
These vertex counterterms are to be inserted in self-energy diagrams with $L<N$ loops
(the ``countertem skeletons''), while the primary skeletons are calculated with the 
renormalized coupling at their vertices.

\vspace{2mm}
\begin{figure}[htb]
\epsfysize=1.5cm
\centerline{\epsffile{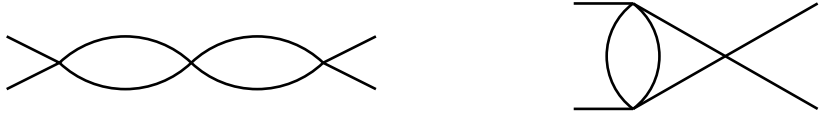}}
\vspace{2mm}
\caption[a]{The   2--loop contribution to $\Lambda$.
\label{3LBPHbis} }
\end{figure}
In the previous examples, we have met
one counterterm of this type, namely the counterterm $\delta\lambda_0$ required to
remove the global divergence of the one-loop contribution to $\Lambda$. 
For the 3-loop primary self-energy in Fig. \ref{4loopPHI}.b, the corresponding, 2-loop, 
contributions to $\Lambda$ are displayed in Fig. \ref{3LBPHbis}, and their various divergences
are exhibited in Fig. \ref{2LLBOX}. These are both one-loop and 2-loop vertex
counterterms, which are included in the vertices of the 2-loop and, respectively, 
one-loop self-energy graphs, as shown in  Fig. \ref{3LCT}. Note that the 
one-loop counterterms, which cancel {\it subdivergences} of $\Lambda$,
involve all the possible channels. By contrast, the
2-loop counterterm, which is a global divergence of this particular
contribution to $\Lambda$, 
involves only the $s$ and $u$ channels (so like $\delta\lambda_0$ in the example of 
Sect. IV).
Clearly, this is related to the irreducibility of $\Lambda$ in one particular channel, the
$t$-channel.

\vspace{2mm}
\begin{figure}[htb]
\epsfysize=4.5cm
\centerline{\epsffile{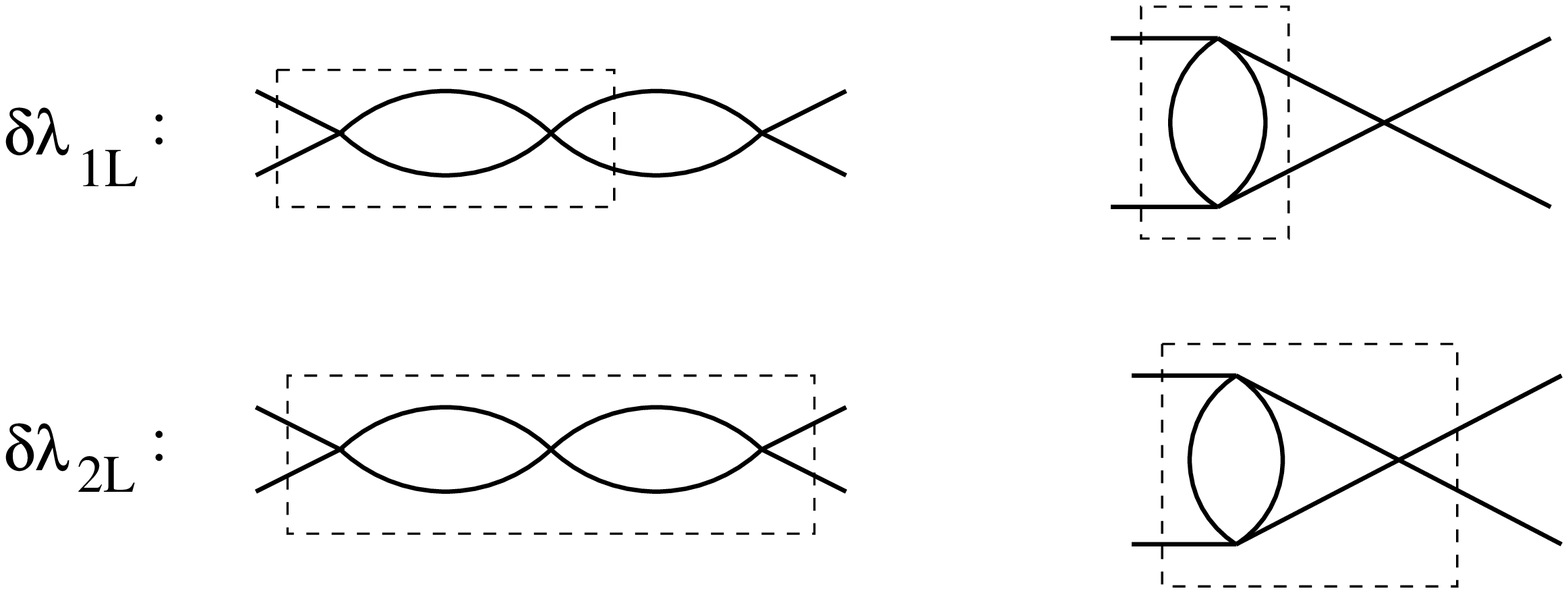}}
\vspace{2mm}
\caption[a]{The subgraphs enclosed in the dashed line boxes correspond to the 
divergences in $\Lambda$, and also to the vertex counterterms shown
in  Fig. \ref{3LCT}. The contribution to $\delta\lambda_{\rm 1L}$ represented
by the diagram on the left corresponds to the $s$ and $u$ channels, while
that on the right corresponds to the $t$ channel.}
\label{2LLBOX} 
\end{figure}

\vspace{2mm}
\begin{figure}[htb]
\epsfysize=3.5cm
\centerline{\epsffile{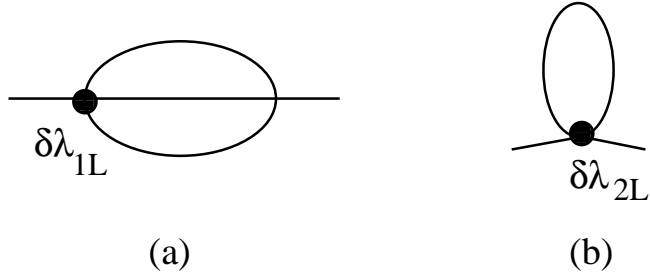}}
\vspace{2mm}
\caption[a]{The counterterm diagrams necessary to remove vertex subdivergences
from the 3-loop self-energy in Fig. \ref{4loopPHI}.b.
\label{3LCT} }
\end{figure}

This example can be generalized to all orders: For a $N$-loop primary self-energy,
the vertex counterterms associated with {\it subdivergences} in $\Lambda$ include all the possible
channels, and appear in the vertices of self-energy skeletons with $2\le L < N-1$ loops
(in such a way that $L+l=N$, where $l$ is the loop number of the vertex counterterm).
As for the {\it global} divergence of $\Lambda$, 
this gives a $(N-1)$--loop counterterm  which
excludes all the topologies which are 2-particle-reducible 
in the $t$ channel. This counterterm is included in the one-loop self-energy
diagram (the ``tadpole'').

 Note that, as a result of this first step, the counterterms needed to renormalize
$\Lambda$ are not explicitly computed, but rather expressed as integrals which involve
the (yet unknown) propagator $D$. However, the divergent parts of these integrals are 
sensitive only to the leading high  momentum behaviour of the propagator, i.e., to $D_2$.
Similarly, $\tilde\Lambda$ (the sum of $\Lambda$ plus the vertex 
counterterms constructed at this step) is a finite functional of $D$, or  $D_2$.

{\it ii\,}) The second step refers to the renormalization and the calculation of the
asymptotically equivalent propagator  $D_2$. To that aim, one proceeds as in the 
3-loop example in Sec. \ref{sec:deltaZ} --- that is, one considers the self-consistent 
gap equation with $m=0$ --- except that, in the general case, one has to include 
in the r.h.s. of the equation for
$\Pi$ all the counterterm diagrams determined in step {\it i\,}). Then, within
dimensional regularisation, the functional $\Pi[D_2]$ (the r.h.s. of the gap
equation) has neither vertex, nor mass, divergences, but only a global field divergence,
 proportional to $K^2$, that can be absorbed in the counterterm
$\delta Z K^2$. After this divergence is subtracted, the ensuing, finite, equation,
can be solved for $\Pi_2$. This gives the propagator $D_2$, which then can be used
to explicitly compute $\tilde\Lambda_2$.

{\it iii\,}) In the third step, the (finite) kernel $\tilde\Lambda_2$ together with the
propagator $D_2$ are used to construct the Bethe-Salpeter equation, together with the
equation for $\delta\Pi/\delta m^2$. By renormalizing these equations, we obtain
 the countertermes $\delta\lambda$ and $\delta m^2$ required to
eliminate the residual vertex subdivergences and, respectively, the mass divergences
from the gap equation. Note that, in contrast to the vertex counterterms for $\Lambda$
(cf. the first step), which involve, at most, $N-1$ loops, those obtained by iterating the
 Bethe-Salpeter equation will receive contributions
with arbitrarily many loops (and similarly for $\delta m^2$).
At this stage, one can use the algebraic derivation in Sect. IV.E
to verify that the gap equation is a finite equation for $\Pi_0\equiv \Pi-\Pi_2$.

The arguments in Sect. IV.E rely on two properties which remain 
true in any order in the skeleton expansion, as we shall see. These properties are:

A) $\Lambda(P,K) - \Lambda(P,0) $, considered as a function of $P$ for fixed $K$, is of degree of divergence
less than zero.

B) The term $\tilde \Pi_r(K)$ in the r.h.s. of 
Eq.~(\ref{tildegapeqn}) is finite and decreases like $1/K^2$ at large
$K$.

The first property, on the behaviour of
$\Lambda(P,K) - \Lambda(P,0) $ at large $P$ for fixed $K$, follows from
the 2-particle irreducibility of $\Lambda$, as we explain now.
The different contributions to the logarithmic behaviour of the 
4-point function at fixed $K$ and large  $P$ are, according to Weinberg's theorem, 
associated 
 to 4-point subgraphs attached to the two 
external lines carrying $P$. Recall that, in Weinberg's terminology, a subgraph is 
any set of lines choosen in the initial graph in such a way that 
there is no vertex attached to only one line of the subgraph 
\cite{Weinberg,Collins:xc}. In 
Fig. \ref{fig:subgraphs}, we give examples of 4-point subgraphs 
attached to $P$ for a graph contributing to $\Gamma(P,K)$. We also 
give an example of a set of lines which is not a subgraph.
\begin{figure}[htb]
\epsfysize=4.cm
\centerline{\epsffile{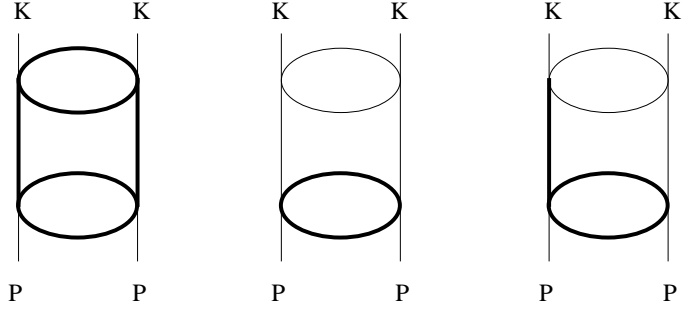}}
\caption[a] {Examples and counter-example of 4-point subgraphs attached to $P$. (a) and (b) are 4-point subgraphs contributing to the dominant behaviour of $\Gamma(P,K)$ at large $P$ and fixed $K$. This leading contribution  is  independent of $K$ in the case of  (a) but the subgraph (b) may give a  $K$-dependent  contribution. (c) is not a subgraph.\label{fig:subgraphs}}
\end{figure}
For both $\Gamma(P,K)$ and $\Lambda(P,K)$, there is always a trivial 
four point subgraph attached to $P$ : the graph itself. Clearly the 
leading behaviour at large $P$ for this contribution is independent of 
$K$, either for $\Gamma(P,K)$ or $\Lambda(P,K)$. The reason is simply 
that for this subgraph, $K$ appears in propagators 
where, by asumption,  the total momenta goes to infinity, while $K$ is kept finite. Of course there 
are other logarithmic contributions. The second diagram in 
Fig. \ref{fig:subgraphs} gives an example of such a 
contribution in which the leading contribution at large $P$ does  depend on $K$. 
But it is easily verified that this particular 2-particle reducible topology
is the only one allowing for a $K$--dependence. Since   $\Lambda(P,K)$ is irreducible,
 its asymptotic behaviour at large $P$ is independent of $K$. Property  A) then follows.

For the second property,  we note that  $\tilde \Pi_r$ is obtained by
 differentiating twice a  skeleton graph for the self-energy.
Thus $\tilde \Pi_r$ has the structure of a 6-point function which is closed
into a two-point function by two $\delta D$, each of degree of divergence $- 4$. 
The internal six-point subgraph itself has degree of divergence $- 2$, and since
all (vertex) subdivergences supposed to be eliminated, this is strictly finite.
When the two effective propagators $\delta D$ are closed on it, the UV finiteness
is preserved because one of the two following situations  occurs: 1) The two
propagators appear together in the same loop; that loop is certainly finite.
2) If a single effective propagator appear in some loop, that loop is certainly
not a tadpole (since the whole graph has the topology of a skeleton), and thus
contains at least one additional propagator belongning to the 
six-point vertex subgraph; that loop is finite too.
Since the overall graph is finite, simple power counting together
with Weinberg's theorem ensures that, at large $K$, $\tilde \Pi_r(K)\sim 1/K^2$.

To conclude, let us briefly discuss the scale dependence of the renormalized 
coupling constant which emerges from the renormalization program described above.
(A similar discussion applies to the $\mu$--dependence of $\Pi_2$ and of the
renormalized mass parameter $m$.)
Consider the $N$--loop approximation to the self-consistent self-energy which
includes {\it all} the primary skeletons with $L\le N$ loops, together with the
corresponding counterterm skeletons. Given the similitude between step
{\it i\,}) above and ordinary perturbation theory, it is clear
that renormalized vertices belonging to diagrams with different numbers of loops
are dressed differently, and thus run differently with $\mu$. The 
respective $\mu$--dependences are non-perturbative, since contributions
of all orders in $\lambda$ are included
(because the corresponding counterterms depend upon $\Pi_2$,
and also because of the iterations performed by the Bethe-Salpeter equation
in the case of the coupling constant which enters the tadpole).
However, when restricted to an expansion in powers of $\lambda$, the 
$\mu$--dependences of the vertices are those expected from the loop order
of the included diagrams. In particular, the couplings in the vertices
of the $N$--loop primary diagrams are not dressed at all, so they are independent
of $\mu$, those which enter the $(N-1)$--loop primary diagrams are running
according to the one-loop beta function, etc. 
The coupling constant in the one-loop self-energy is exceptional, in that
it exhibits the most developped running
(perturbatively correct up to loop order $N-1$).

The one-loop example discussed in Sect. III (see, especially, Sect. III.D)
suggests that it should be possible to exploit the non-perturbative character of the
$\mu$--dependence of the vertices
 in order to minimize the scale dependence of the self-consistent solutions
for $\Gamma$ and $\Pi$. In that one-loop case, we have in fact been able to
verify explicitely that both $\Gamma$ and $\Pi$ are
strictly independent of $\mu$. In the general case, the corresponding analysis
is considerably more involved, and will not be pursued here. We shall only
emphasize a specific feature of the $\Phi$--derivable approximations,
which is that vertices in different diagrams run differently with $\mu$.
This should be taken into account when trying to reduce (and possibly eliminate) 
the scale dependence of the physical observables.


\newpage
    
\begin{thebibliography}{99}

\bibitem{Luttinger:1960}
J.~M. Luttinger and J.~C. Ward, {\it Phys. Rev.\/} {\bf 118}, 1417 (1960).

\bibitem{Baym} G. Baym and L. Kadanoff, Phys. Rev. {\bf 127} 22 
(1962); G. Baym, Phys. Rev. {\bf 127}, 1391
(1962).

\bibitem{DeDominicis:1964}
C.~{De Dominicis} and P.~C. Martin, {\it J. Math. Phys.\/} {\bf 5}, 14, 31
  (1964).

\bibitem{Cornwall:vz}
J.~M.~Cornwall, R.~Jackiw and E.~Tomboulis,
Phys.\ Rev.\ D {\bf 10}, 2428 (1974).


\bibitem{Blaizot:1999ip}
J.~P. Blaizot, E.~Iancu and A.~Rebhan,   Phys. Rev. Lett.\ {\bf 83}, 2906 (1999); 
 Phys. Lett.\ {\bf B470}, 181 (1999); Phys.\ Rev.\ D {\bf 63}, 065003 (2001).

\bibitem{Vanderheyden98}
B. Vanderheyden and G. Baym, J. Stat. Phys. {\bf 93}, 843 (1998).

\bibitem{Peshier:2000hx}
A.~Peshier, 
  Phys. Rev. {\bf D63}, 105004 (2001).

\bibitem{Braaten:2001en}
E.~Braaten and E.~Petitgirard, Phys. Rev.  {\bf D65},
  041701 (2002); Phys. Rev.  {\bf
  D65}, 085039 (2002).

\bibitem{Karsch:1997gj}
F.~Karsch, A.~Patk{\'o}s and P.~Petreczky, Phys. Lett.  {\bf B401}, 69 (1997).


\bibitem{Aarts:2002dj}
J.~Berges,  Nucl. Phys. {\bf A699}, 847 (2002);
G.~Aarts, D.~Ahrensmeier, R.~Baier, J.~Berges and J.~Serreau,
Phys.\ Rev.\ D {\bf 66}, 045008 (2002).

\bibitem{vanHees:2002bv}
H.~van Hees and J.~Knoll,
Phys.\ Rev.\ D {\bf 66}, 025028 (2002)

\bibitem{Arrizabalaga:2002hn}
A.~Arrizabalaga and J.~Smit,
Phys.\ Rev.\ D {\bf 66}, 065014 (2002)


\bibitem{Dolan:qd}
L.~Dolan and R.~Jackiw,
Phys.\ Rev.\ D {\bf 9}, 3320 (1974).


\bibitem{Drummond:1997cw}
I.~T.~Drummond, R.~R.~Horgan, P.~V.~Landshoff and A.~Rebhan,
Nucl.\ Phys.\ B {\bf 524}, 579 (1998)

\bibitem{Collins:xc}
J.~C.~Collins,
``Renormalization. An Introduction To Renormalization, The 
Renormalization Group, And The Operator Product Expansion,''
Cambridge, Uk: Univ. Pr. ( 1984) 380p


\bibitem{Blaizot:2002nh}
J.~P.~Blaizot, R.~Mendez-Galain and N.~Wschebor,
Annals Phys.\  {\bf 307}, 209 (2003).



\bibitem{vanHees:2001ik}
H.~van Hees and J.~Knoll,
Phys.\ Rev.\ D {\bf 65}, 025010 (2002)


\bibitem{VanHees:2001pf}
H.~Van Hees and J.~Knoll,
Phys.\ Rev.\ D {\bf 65}, 105005 (2002)

\bibitem{BPH}
N.N. Bogoliubov and O.S. Parasiuk, Acta Math. {\bf 97}, 227 (1957);
K. Hepp, Commun. Math. Phys. {\bf 2}, 301 (1966).

\bibitem{Blaizot:2003br}
J.~P.~Blaizot, E.~Iancu and U.~Reinosa,
Phys.\ Lett.\ B {\bf 568}, 160 (2003).



\bibitem{Kap:FTFT}
J.~I. Kapusta, {\it Finite-temperature field theory\/} (Cambridge University
  Press, Cambridge, 1989).

\bibitem{LeB:TFT}
M.~{Le Bellac}, {\it Thermal Field Theory\/} (Cambridge University Press,
  Cambridge, 1996).



\bibitem{BR86} J.P.~Blaizot and G.~Ripka, {\em Quantum Theory
of Finite Systems} (MIT Press, Cambridge, 1986).

\bibitem{Weinberg}
S. Weinberg, Phys.\ Rev. {\bf 118},  838 (1960).

\bibitem{Ramond:yd}
P.~Ramond,
{\em Field Theory: A Modern Primer} (Addison-Wesley, 1989).



------------------------------------------------------------------------



\end{thebibliography}
\end{document}